\documentclass[aps,twocolumn,pre,showpacs,superscriptaddress,amsmath,amssymb]{revtex4-1}
\usepackage{amsmath,comment}
\usepackage{amssymb,graphicx}
\usepackage{color,txfonts}

\definecolor{dkgreen}{rgb}{0,0.6,0}
\definecolor{gray}{rgb}{0.5,0.5,0.5}
\definecolor{mauve}{rgb}{0.58,0,0.82}

\begin{document}

\title{Social contagion with degree-dependent thresholds}

\author{Eun Lee}
\affiliation{Department of Energy Science, 
Sungkyunkwan University, Suwon 16419, Korea}

\author{Petter Holme}
\affiliation{Department of Energy Science, 
Sungkyunkwan University, Suwon 16419, Korea}

\begin{abstract}
We investigate opinion spreading by a threshold model in a situation where the influence of people is heterogeneously distributed. We focus on the response of the average opinion as a function between the trend between out-degree (number of neighbors)---effectively the strength of influence of a node---and the threshold for adopting a new product or opinion. We find that if the coupling is very positive, the final state of the system will be a mix of different opinions otherwise it will it converges to a consensus state. We find that this cannot be simply explained as a phase transition, but emerges from a combination of mechanisms and their relative dominance in different regions of parameter space.
\end{abstract}

\maketitle

\section{Introduction}
\label{sec:introduction}
In everyday life, people  make decisions about purchases, political opinions, where to go, and so on. However, even though each decision is made by a person, this person is not isolated, but constantly affected by others. The process of ideas spreading over social networks is well established in the social sciences~\cite{rogers,granovetter,schelling,axelrod} and arguably influenced by the structure of the network. As an early example, Granovetter suggested that ``the number or proportion of others who must make one decision before a given actor does so'' is a fundamental personal trait driving the contagion~\cite{granovetter}. It is straightforward to see that this type of dynamics is sensitive to network structure~\cite{watts,handjani,valente,dodds,melnik,gomez}. This paper investigates what happens if a threshold, such as the one Granovetter mentions, depends on the how influential a node is. We use the out-degree---the number of people being influenced by the agent---as a measure of the power to influence others. We tune the correlation between the out-degree and the threshold (keeping the average threshold level constant), to investigate how  a correlation between the threshold and the influence determines the opinion dynamics. There are reasons to believe such a correlation exists. In particular, it is easy to imagine influential agents to be celebrities whose publicly visible choices are carefully made, with the impact to the population in consideration. Therefore we would, in real systems, expect a positive correlation between threshold and influence; i.e., influential actors to be more conservative. An opposite effect is also conceivable in situations where products are marketed via influential actors. In the absence of data, we scan the entire parameter space from the maximum negative to the maximum positive correlations.

Going back to the threshold-model literature. Since Granovetter, there has been many papers investigating the relation between the network structure and opinion dynamics~\cite{watts,handjani,valente,dodds,melnik,gomez}. Perhaps the most influential paper was Watts's study of threshold models as an explanation of cascades or fads~\cite{watts}.
Other papers focused on how the seed nodes' network position influence the subsequent spreading process~\cite{singh,gleeson2007}. Ref.~\cite{dodds2} studied the influence of degree-degree correlations. Refs.~\cite{whitney,gleeson2008} studied the role of clustering. Refs.~\cite{osman,charles,kyumin} studied threshold models on multiplex networks, and Refs.~\cite{karimi_holme,takaguchi_masuda_holme} studied them on temporal networks. In this body of literature, most works, in contrary to Granovetter's original idea, use a homogeneous threshold. For this reason, the effect of heterogeneous thresholds is still not fully understood. Even though there are some exceptions like Refs.~\cite{wang,marton} that studied the effect of heterogeneous threshold, however, the heterogeneity in ~\cite{wang} still has constraints to classify the threshold in two types and the threshold in ~\cite{marton} did not include the correlation case. In the related problem of disease spreading, the corresponding situation of a the heterogeneous susceptibility has been studied in e.g.\ Refs.~\cite{hidalgo, miller}. At a basic level, there must be an individual variation also in threshold. Probably there are other factors that also affect a decision, but introducing a heterogeneity in the threshold is a reasonable first extension of the uniform threshold model.

As for the influence, it is fair to assume it is heavy tailed. E.g.\ Ref.~\cite{moon} argues that it is power-law distributed. Many other characteristics that one would assume correlated with influence (wealth, follower counts, etc.)\ also show a skewed, sometimes power-law distribution~\cite{mejn:powerlaw}. For these reasons, we model the out-degree as power-law distributed. We keep the in-degree, on the other hand, fixed. This is motivated by the Dunbar number---that people (for cognitive reasons) have a limited capacity of the number of concurrent friendships~\cite{galesic,dunbar}. This suggests the number of people one are directly influenced by is also limited. Even if it probably also varies, it would be more narrowly distributed than the out-degree, so for simplicity we make it constant.

In the remainder of the paper, we will present the model in detail and go through the numerical results.

\section{Model}
\label{sec:model}

As mentioned, we construct a model to mimic a networked society where the power to influence (out-degree) others is broadly distributed, but not the number of people one is influenced by. Furthermore, we assume there can be a coupling between threshold $\phi_i$ and the power to influence (i.e., the out-degree) $k_i$, and implement it by setting it deterministically based on $k_i$.

Specifically, we start by assigning desired out-degrees $x_i$ to $N$ isolated nodes from a truncated power-law distribution.
\begin{equation}
p(x) = \left\{ \begin{array}{ll}
\sim x^{-\gamma} & \mbox{for $x\in [1,N^{1/(\gamma - 1)}]$}\\
0 & \mbox{otherwise}
\end{array}\right. .
\end{equation}
As $N\rightarrow\infty$ this converges to a power-law, but the truncation makes dampens the fluctuations and make the convergence of the simulation quicker. Then we proceed, like the configuration model~\cite{newman:book}, to add links by randomly choosing pairs $(i,j)$ of nodes and add a directed edge from $i$ to $j$ if:
\begin{enumerate}
\item $i\neq j$.
\item $k_i < x_i$.
\item $q_j < y$, where $q_i$ is the in-degree of $j$ and $y$ is the maximal in-degree.
\end{enumerate}
This process is iterated until there are less than two nodes with the desired degree conditions unsatisfied. This guarantees that there are no self-edges and all the out-degree distribution is power-law in the large-$N$ limit, and the in-degrees are constant (with the possible exception of one node).

\begin{figure}
\includegraphics[width=0.9\linewidth]{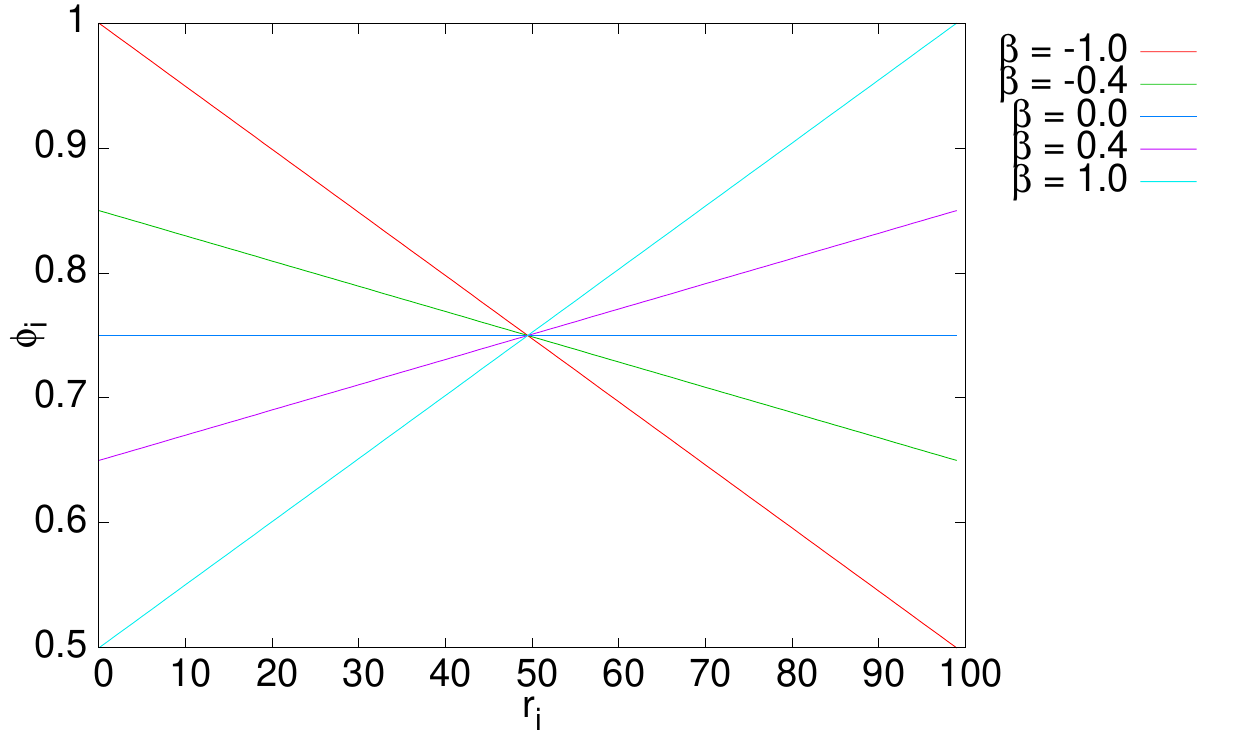} 
\caption{(Color online) Threshold $\phi_i$ as a function of sorted out-degree rank $r_i$ with different coupling parameter $\beta$.}
\label{fig:threshold}
\end{figure}

Then we proceed to assign thresholds based on the out-degrees. To be able to tune the relation between the threshold and out-degree, we need some conditions to be fulfilled. Fist, we have to keep the average threshold independent of any control parameter. Second, the threshold should be strictly larger than $1/2$ since we assume the social influence is positive. To meet those prerequisites, we use the following formula
\begin{equation}
\phi_i = \frac{\beta}{2} \frac{r_i}{N-1} + \frac{3-\beta}{4} 
\label{threshold}
\end{equation}
Here, $\beta$ is the variable for controlling how strongly the threshold is affected by the degree. The sign of $\beta$ also determines the parity of this influence---if it is positive an influential node has a high threshold, if it is negative, the threshold is lower the more influential the node is. $r_i$ is the roughly speaking the rank of $i$ (with $0$ being the node of lowest $k_i$ and $N-1$ being the highest) with respect to the out-degree---we rank nodes with the same degree randomly. Fig.~\ref{fig:threshold} illustrates how $\beta$ controls the relation between $\phi_i$ and $r_i$.

\begin{figure}
\includegraphics[width=0.7\linewidth]{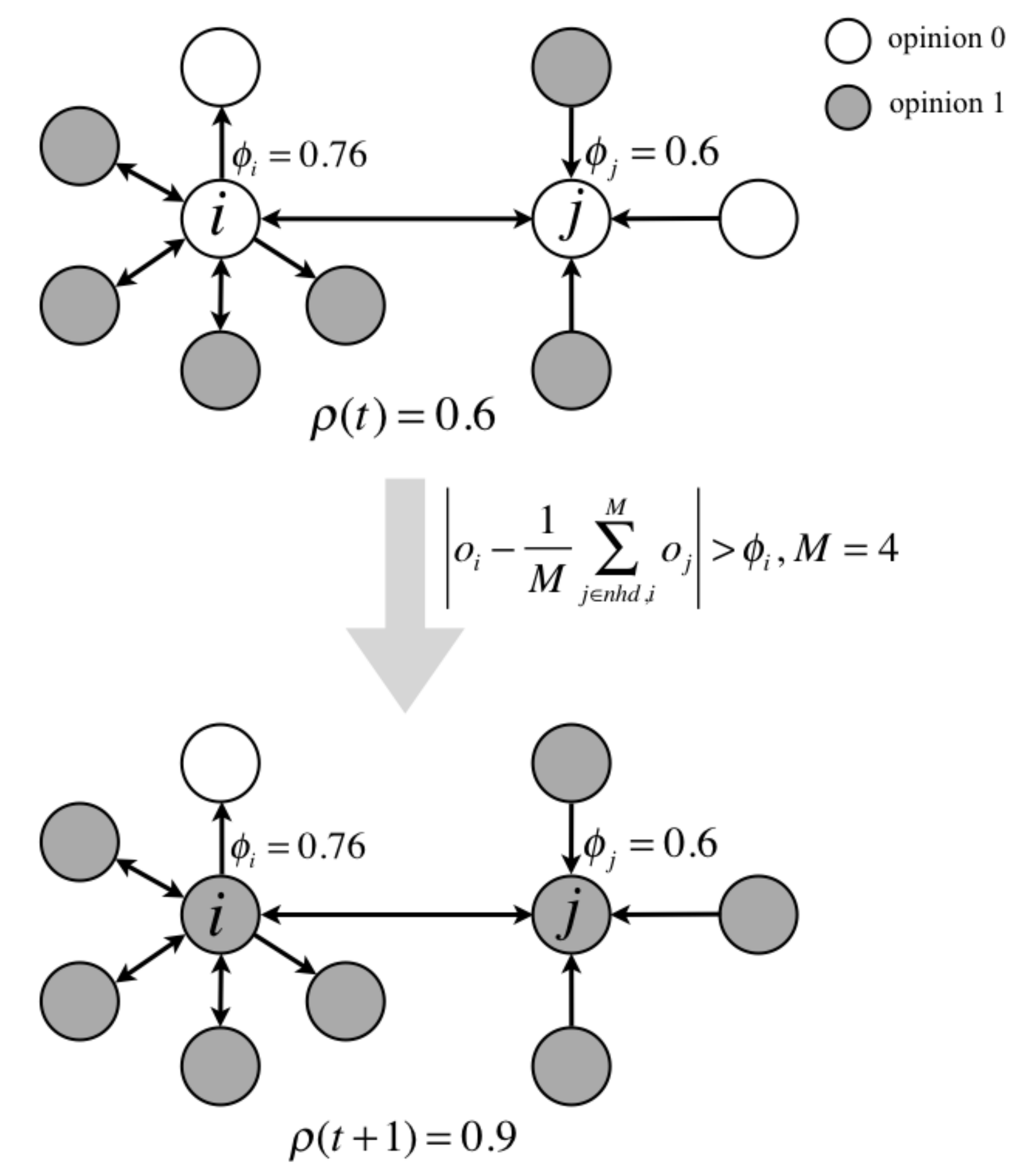} \\
\caption{(Color online) Schematic description of the updating rule. $\phi_i$ and $\phi_j$ are thresholds of nodes $i$ and $j$. $\rho(t)$ is the averaged opinion of the system at time $t$ (measured in Monte Carlo simulation steps), and $\rho(t+1)$ is the averaged opinion after the first time step. Here, $M = 4$ for simplicity. The arrow is indicating the direction of influence. After the comparison process of the opinions in the extended neighborhood of $o_i$, node $i$ changes its opinion $0$ (white) to the opinion $1$ (grey).} 
\label{fig:model}
\end{figure}

With the network constructed and the thresholds assigned, we run simulations of opinion spreading using a threshold-type dynamics in the spirit of Granovetter~\cite{granovetter}. This is illustrated in Fig.~\ref{fig:model}. We initialize the binary opinion of all nodes randomly $o_i\in\{0,1\}$ with a probability $p$ for $o_i=1$ and $1-p$ for $o_i=0$. $p$ is meant to reflect the probability of an outcome if all the population has the full information on the subject. In this paper we will use $p=0.6$. Then we proceed updating the states by:
\begin{enumerate}
\item Picking a node $i$ at random.
\item Calculating the average opinion $\bar{o}_i$ of its in-neigh\-bor\-hood (i.e.\ the nodes $i$ can be influenced by).
\item If $\bar{o}_i-o_i>\phi_i$ change $o_i$ from $0$ to $1$; else if $\bar{o}_i-o_i<-\phi_i$ change $o_i$ from $1$ to $0$.
\end{enumerate}

\begin{figure*}
\begin{tabular}{ll}
(a) & (b)\\
\includegraphics[width=0.45\textwidth]{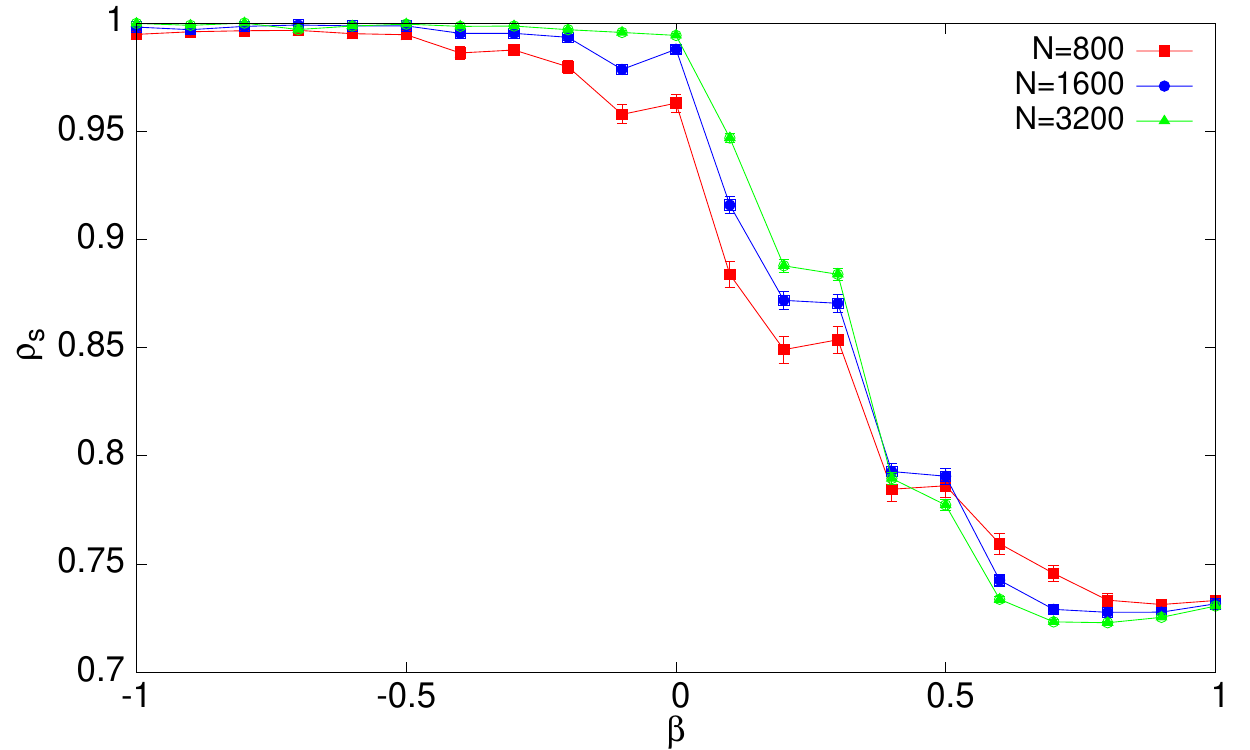} &\includegraphics[width=0.45\textwidth]{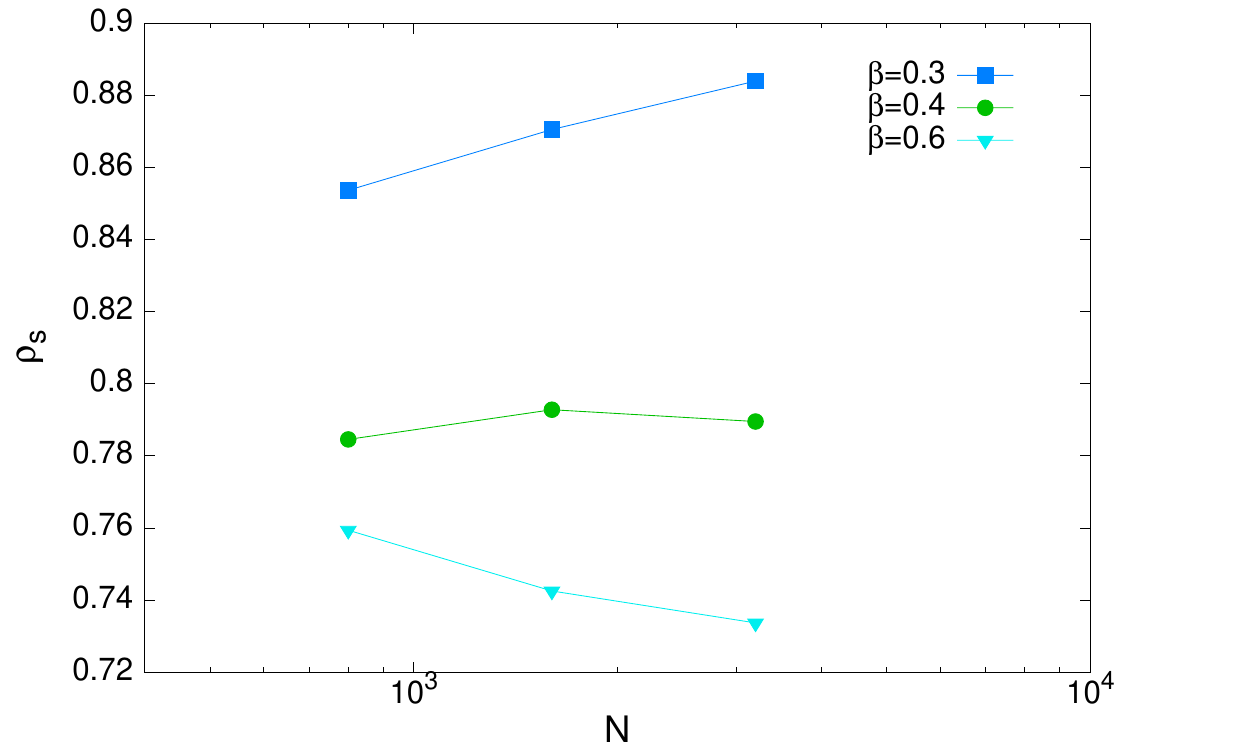} \\
\end{tabular}
\caption{(Color online) Average opinion in steady-state ($\rho_s$) with (a) different $\beta$ for $N=800$, $1600$, $3200$, and (b) average opinion in steady state ($\rho_s$) with different sizes for the transition behavior at $\beta = 0.4$ ($n=2000$, standard error is shown). }
\label{fig:size}
\end{figure*}

\section{Results}

The model inevitably reaches a fixed state where no agent change its opinion. As our first numerical analysis, we measure the average opinion \begin{equation}\rho(t) = \frac{1}{N} \sum_{i=0}^{N-1} o_i(t)\end{equation} at this final, fixed state ($\rho_s$). Fig.~\ref{fig:size}(a) displays $\rho_s$ as a function of $\beta$ for system sizes ranging from $N=800$ to $N=3200$ (we averaged it $2000$ times, $n=2000$). In general, $\rho_s$ decreases with the coupling strength $\beta$, and there is a clear asymmetry between negative coupling and positive coupling. As for negative $\beta$, the system is likely to reach a majority consensus state (i.e.\ the opinion $1$, because of the initial condition is $p=0.6>1/2$). For a positive coupling, the final state has $\rho_s$ between $p$ and $1$ for all sizes and $\beta$ values.

The most notable result of Fig.~\ref{fig:size} is perhaps that for $\beta\simeq 0.4$ all the curves decreases with size, while for $\beta < 0.4$ they increase. This could mean that $\rho_s=1$ for all $\beta < 0.4$ in the $N\rightarrow\infty$ limit, and $\rho_s<1$ otherwise. I.e., that there is a phase transition between a phase where the system can integrate all the information (small $\beta$) and a phase where the opinion, or thing to adopt, would not spread to the entire population. This is however hard to conclusively confirm because of the large fluctuations makes the convergence prohibitively slow (thus stopping us from scaling up the size). This common in network because of the short path lengths (scaling logarithmically or even slower~\cite{chunglu}). In addition, there are step-like structures. Such are common in threshold models on networks~\cite{watts} especially those with fat-tailed degree distributions. The reason for the steps is that since there are many low-degree nodes, then when $\beta$ passes a multiple of one over their degree, the dynamics can change quite dramatically. So for the reasons mentioned, we cannot present any stronger evidence for a phase transitions than the different trends in $\rho_s$ as a function of $N$ at $\beta\approx 0.4$. This trend is plotted directly in Fig.~\ref{fig:size}(b).

\begin{figure*}
\begin{tabular}{lll}
(a) & (b) & (c)\\
\includegraphics[width=0.33\linewidth]{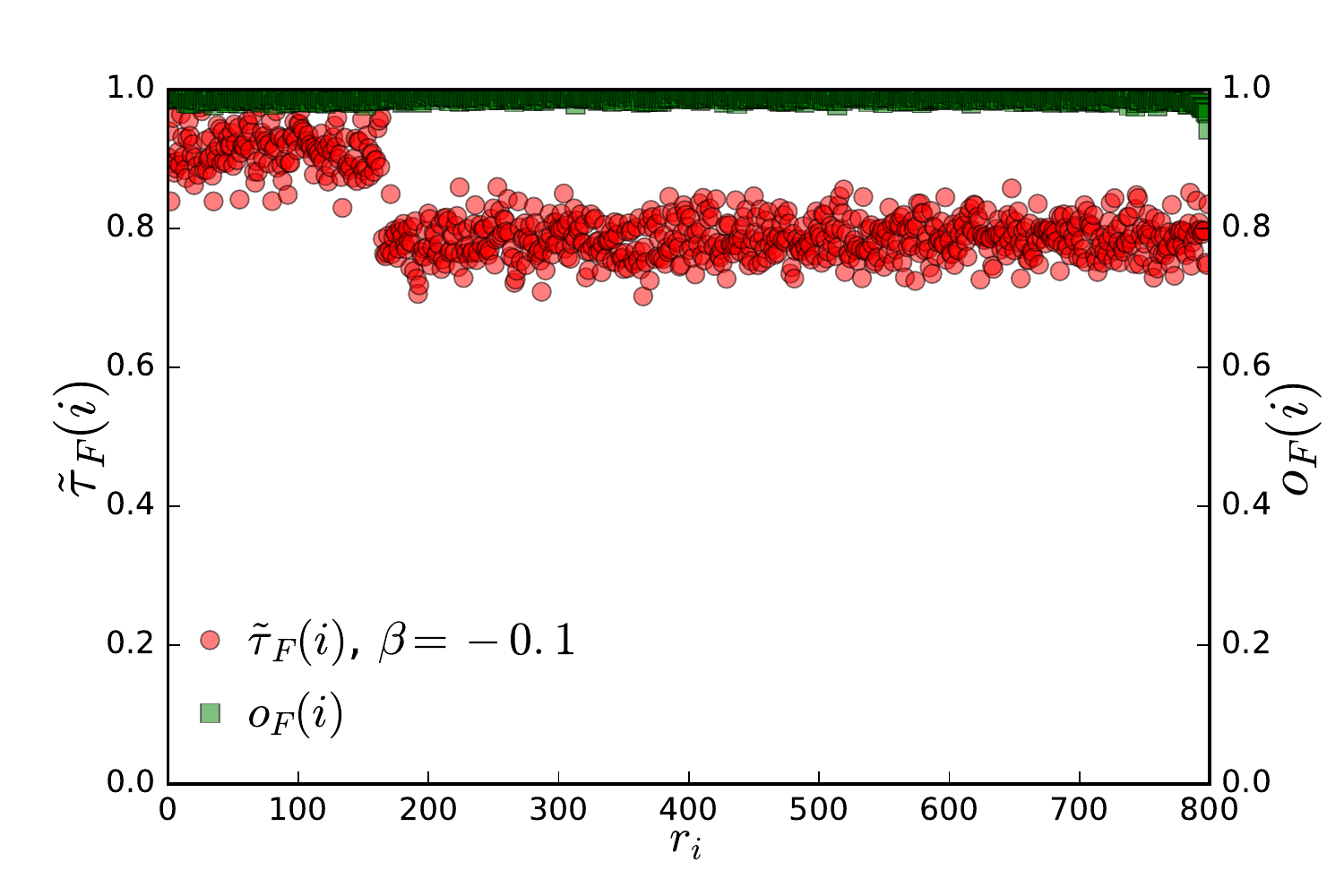} &
\includegraphics[width=0.33\linewidth]{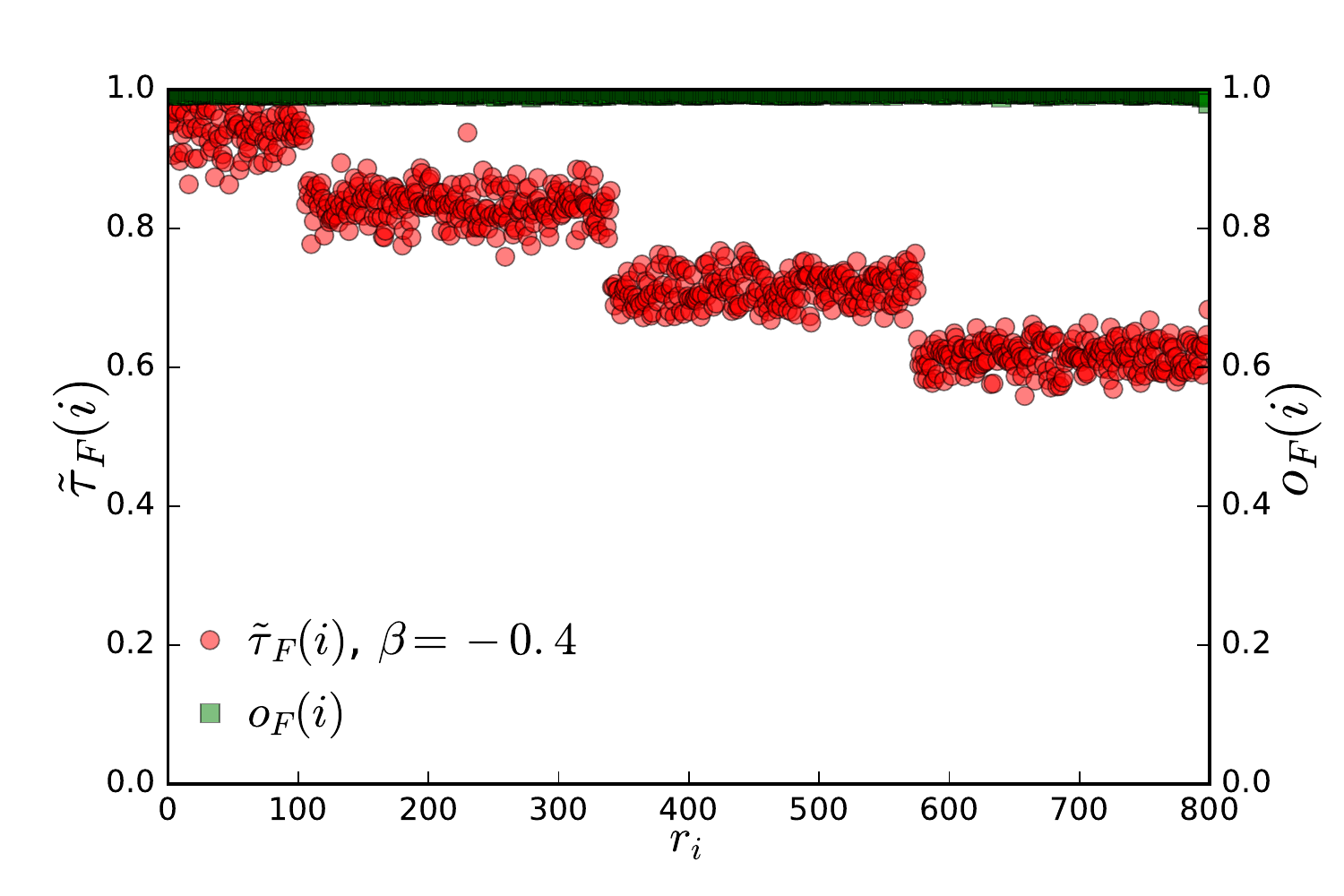}&
\includegraphics[width=0.33\linewidth]{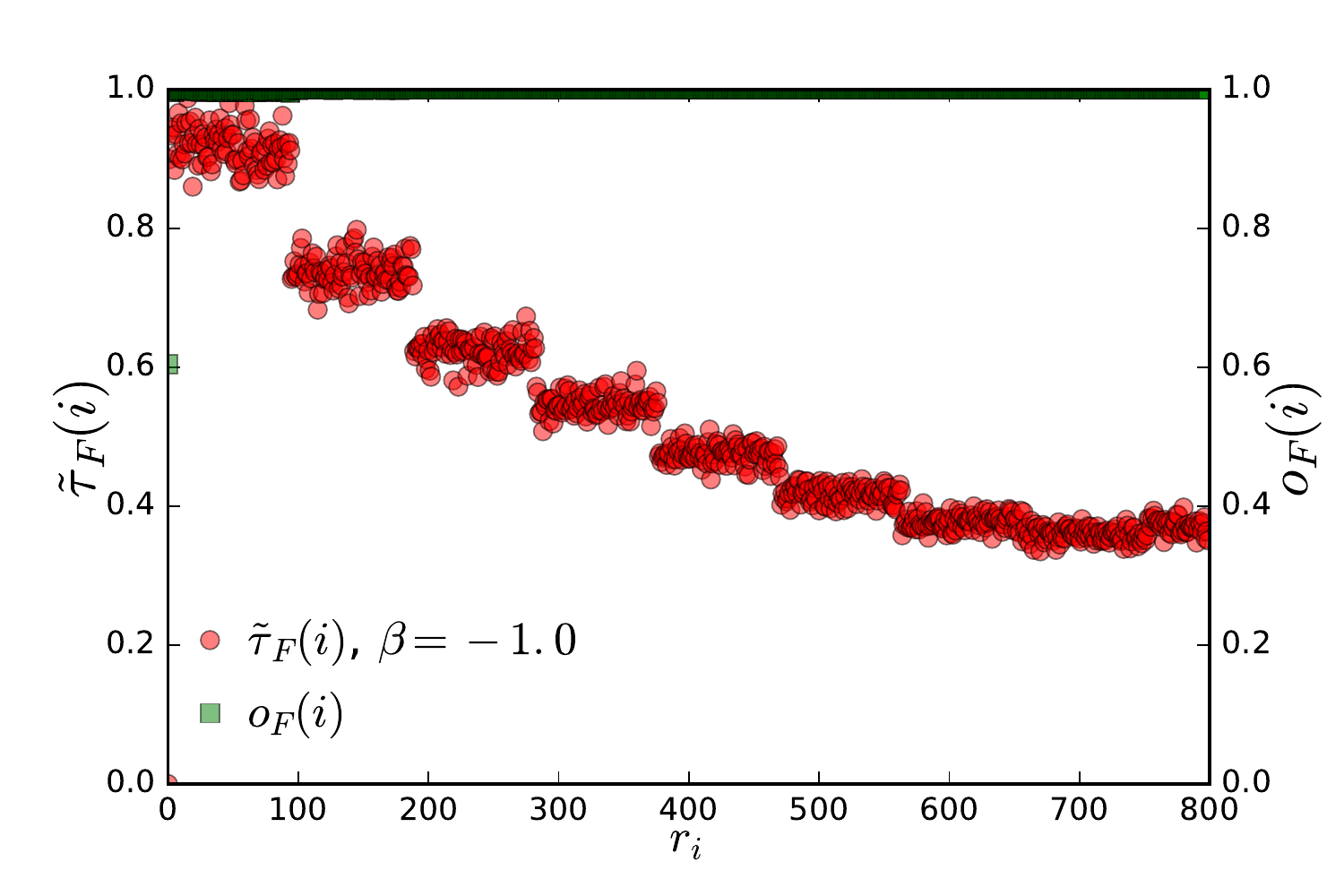}\\
\end{tabular}

\begin{tabular}{lll}
(d) & (e) & (f) \\
\includegraphics[width=0.33\linewidth]{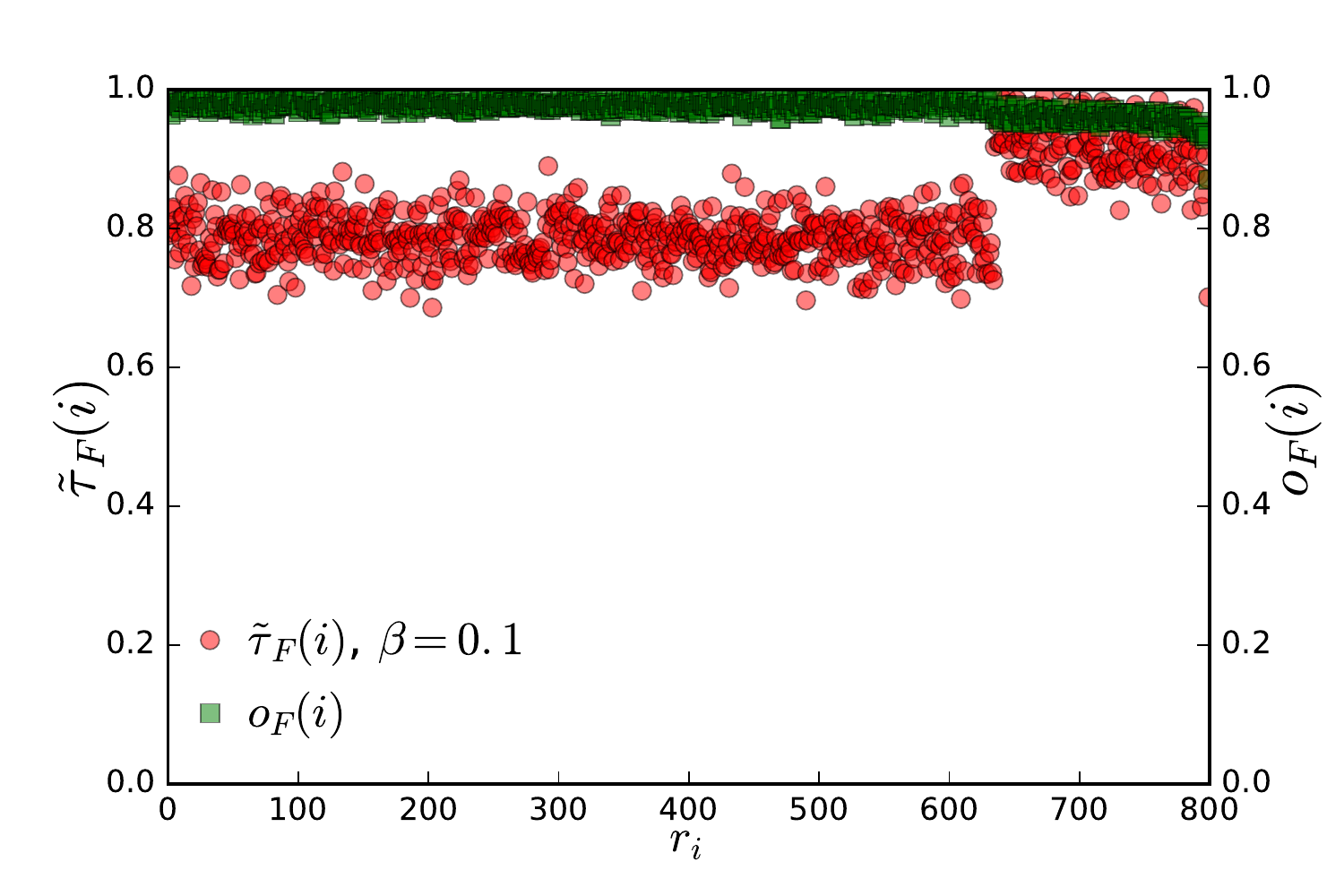} &
\includegraphics[width=0.33\linewidth]{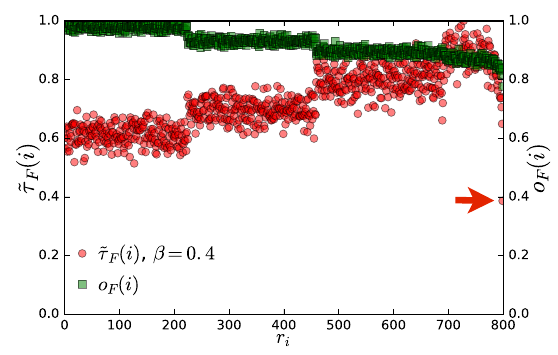}&
\includegraphics[width=0.33\linewidth]{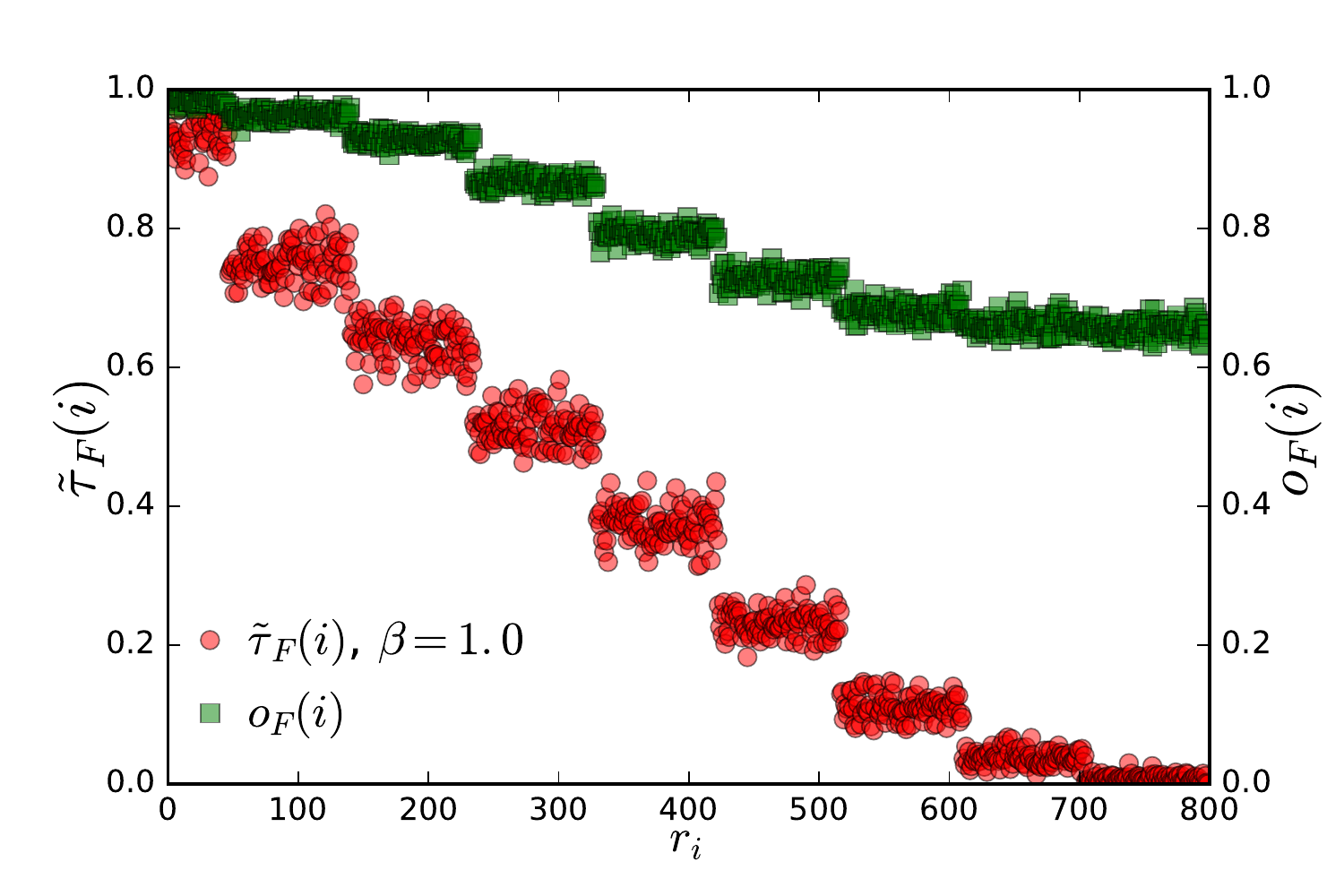}\\
\end{tabular}

\caption{(Color online) Normalized fixation time ($\tilde{\tau}_F(i)$) and finalized opinion ($o_F(i)$) for  different coupling strengths $\beta$ as functions of the rank $r$ of the nodes. Panel (a) shows $\beta = -1$, (b) $\beta = -0.4$, (c) $\beta = -0.1$, (d) $\beta = 0.1$, (e) $\beta = 0.4$ and (f) $\beta = 1$. The arrow in panel (e) points to the most influential node that becomes fixed first. Each $\tilde{\tau}_F(i)$ and $o_F(i)$ averaged for $2000$ runs ($n=2000$) with $N=800$. }
\label{fig:fixation}
\end{figure*}

Another illuminating quantity is the fixation time $\tau_F(i)$ of a node $i$---the time until no more change of the opinions will ever happen. We denote $i$'s averaged final opinion by $o_F(i)$ with $n$ ensembles, so $o_F(i)$ could have a float even though $o_i$ itself is only possible $0$ or $1$.  
To see the relative fixation time of a node, we normalized $\tau_F(i)$ by $t_s$, which is the largest fixation time for a dynamics. We plot the results for normalized fixation time $\tilde{\tau}_F(i)$ ($\tau_F(i)/t_s$) and $o_F(i)$ as functions of the rank of the out-degree of $i$ in Fig.~\ref{fig:fixation}. 
Just like Fig.~\ref{fig:size}, the step-like shapes of Fig.~\ref{fig:fixation} obviously come from the fact that degrees are integers. 

A change of an opinion can occur when
\begin{equation}|\bar{o}_i-o_i| > \phi_i .\label{eq:oioi}\end{equation}
For the change from $0$ to $1$ of a node $i$, $\bar o_i > \phi_i$ should be satisfied. As $\bar o_i$ is the average opinion of $i$'s neighbors, it corresponds with $m_i/M$ (here, $m_i$ is the number of $i$'s neighbors who are having opinion $1$). After we plug $m_i/M$ into Eq.~\ref{eq:oioi}, we see that $m_i/M > \phi_i$. From this, the minimum $m_i$  satisfying the condition is $m_i=M\phi_i$. Consider, for example, $\beta = -0.1$  (Fig.~\ref{fig:fixation}(a)). In this case, the range of threshold values are $\phi_i \in [0.725, 0.775]$. Even more specifically, consider $i=164$ and $165$. $\phi_{164} = 0.76474$ and $\phi_{165} = 0.76468$ to clarify the origin of the border between groups. For the concerned nodes, $m_i = M\phi_i$ gives $m_{164} = 13.0005\dots$ and $m_{165} = 12.9995\dots$. It means node $164$ needs at least $14$ neighbors having an opposite opinion to change it (since in-degree is an integer). The same number for $165$ is $13$. Now, we can simplify the range of possible $m_i$ for the change from $0$ to $1$ as
\begin{equation}
m_i \geq \lceil M\phi_i \rceil,
\label{0to1}
\end{equation}
where $\lceil\; \dot\; \rceil$ denotes rounding to the closest larger integer.

Since the condition for the change from $1$ to $0$ is $m_i < M(1-\phi_i)$, the range of possible $m_i$ for the opinion change can be written as 
\begin{equation}
m_i \leq \lfloor M(1-\phi_i) \rfloor.
\label{1to0}
\end{equation}
where $\lceil\; \dot\; \rceil$ is the operation to round to the closest smaller integer.

For both changes---$0$ to $1$ or $1$ to $0$---nodes in each group divided by a border sharing the same range of $m_i$. In the example above, nodes $i \leq 164$ comprise group with relatively large $\tilde{\tau}_F(i)$. Nodes $i > 164$ are composing the second group having relatively small $\tilde{\tau}_F(i)$ (Fig.~\ref{fig:fixation}(a)). As nodes in each group share simultaneous fixation for $\tilde{\tau}_F(i)$ and $o_F(i)$, we are going to use the concept of a group from now on. The index ($q$) will be started from the lower ranked group ($G_q(\beta)$). In the example, there are two groups: $G_0(\beta=-0.1)$ for $i \in [0,164]$ and $G_1(\beta=-0.1)$ for $i \in [165, 799]$ when $\beta = -0.1$. In particular, we define the initiator group who accepts opinion $1$ from the starting point as $\hat{G}(\beta)$ for each $\beta$. In the example of $\beta=-0.1$, $G_1(\beta=-0.1) = \hat{G}(\beta=-0.1)$.

Figures~\ref{fig:fixation}(a), (b) and (c) display $\tilde{\tau}_F(i)$ and $o_F(i)$ for $\beta < 0$. The most conspicuous feature is the early fixation of influential nodes having low threshold, and it is consistent regardless of a strength. Most of nodes in $\hat{G}(\beta<0)$ fixed their opinion as $1$ ($o_F(i) = 1$ for $i \in \hat{G}(\beta<0)$) in very early stage. Afterwards, nodes in other groups sequentially fix the final opinion around $1$ ($o_F(i) \sim 1$). It is because initiator group ($\hat{G}(\beta<0)$) has high out-degree, so it could easily change the opinions of neighbors. This can then trigger an avalanche that spreads fast through the system. When we consider the fact that nodes having low threshold generally vary more and take longer time to set their opinion by neighbor's opinion, Figs.~\ref{fig:fixation}(a), (b), and (c) are  showing how the coupling between influence level (out-degree) and threshold could create a rather unexpected dynamics.

If $\beta=0.1$, most of nodes involving in $\hat{G}(\beta=0.1)$ fix their opinion to $1$. Given the order of fixation (Fig.~\ref{fig:fixation}(d)), the mechanisms looks similar to when $\beta<0$. When $\beta=0.4$ (Fig.~\ref{fig:fixation}(e)), the situation is clearly different. The convergence to $1$ is fast from all nodes, but becomes less common for $r_i>0.9N$.  $G_{2}(\beta=0.4)$  has a yet lower $o_F(i)\sim 0.85$ compared to $G_{1}(\beta=0.4)$. Even $o_F(i)$ of the highest threshold group ($G_{3}(\beta=0.4)$) starts decreasing around $r_i=0.8N$ and its $\tilde{\tau}_F(i)$ is similar with that of $G_{2}(\beta=0.4)$. The arrow in Fig.~\ref{fig:fixation}(e) points out the highest ranked node (with the highest threshold). This node finalizes its opinion earlier than $\tilde{\tau}_F(i)$ of $\hat{G}(\beta=0.4)$. Thus, the highest influential node has not had time to influence others by the time $\hat{G}(\beta=0.4)$ fixes its opinion. From $\beta=0.4$, the fixation order of dynamics is reversed---from more to less influential nodes. Also $o_F(i)$ of each group is divided into diverse ranges. Even the minimum $o_F(i)$ is around $0.6$, which is similar to the initial probability of opinion $1$ for the highest ranked nodes and the maximum $o_F(i)$ is close to $1$ even though $\hat{G}(\beta=1.0)$ takes longest to fix its opinion. 
    
To understand the dynamics for $\beta\geq 0.4$, assume $P_i(1 \rightarrow 0)$ ($P_i(0 \rightarrow 1)$) is the probability of opinion change from $1$ to $0$ ($0$ to $1$) at the first time step of simulation.
 For $P(1 \rightarrow 0)$, as described above, $1-\bar{o}_i > \phi_i$ need to be satisfied ($\bar{o}_i$ being the average opinion of $i$'s neighbors). Then, we can rewrite the rule as $1-m_i/M > \phi_i$ since $\bar{o}_i$ equals $m_i/M$ ($m_i$ is the number of $i$'s neighbors having opinion $1$ and $M$ is the sum of in-degree for all nodes). This condition can be rewritten as  $\lfloor M(1-\phi_i) \rfloor \geq m_i$ since $m_i$ should be an integer. For each node $i$, $P_i(1 \rightarrow 0)$ should be a sum of the probability for contagion for all the possible $m_i$ ($m_i \in [0,\dots,\lfloor M(1-\phi_i) \rfloor]$). Then, $P(1 \rightarrow 0)$ will be
\begin{equation}
P(1 \rightarrow 0) = \sum_{m_i=0}^{\lfloor M(1-\phi_i) \rfloor} P(m_i).
\label{onetozero}
\end{equation}
In the same way, for $P(0 \rightarrow 1)$  we obtain 
\begin{equation}
P(0 \rightarrow 1) = \sum_{m_i = \lceil M\phi_i \rceil
}^{M} P(m_i).
\label{zerotoone}
\end{equation}

Thanks to the initial random distribution of opinions, we can know that the distribution of $m_i$ as a binomial distribution that can be expressed by $P(m_i) = B(M,m_i,p) = \binom {M}{m_i}p^{m_i}(1-p)^{M-m_i}$. By using $P(m_i)$, the master equation for $o_i(t)$ at next time step could be written as
\begin{equation}
o_i(t+1) = o_i(t) - pP(1 \rightarrow 0) + (1-p)P(0 \rightarrow 1).
\label{oi}
\end{equation}
By plugging Eq.~\eqref{onetozero} and Eq.~\eqref{zerotoone} into Eq.~\eqref{oi}, it can be rewritten as
\begin{equation}
o_i(t+1) = o_i(t) - p\sum_{m_i=0}^{\lfloor M(1-\phi_i) \rfloor} P(m_i) + (1-p)\sum_{m_i = \lceil M(\phi_i) \rceil
}^{M} P(m_i).
\label{newoi}
\end{equation}
This equation is, unfortunately, hard to analyze analytically. This is much because the threshold is based on the rank rather than the actual out-degree. In future works, we plan to relax the condition that the average threshold should be conserved to be able to make headway in such a calculation.
 
\begin{figure}
\begin{tabular}{l}
(a)\\
\includegraphics[width=0.94\linewidth]{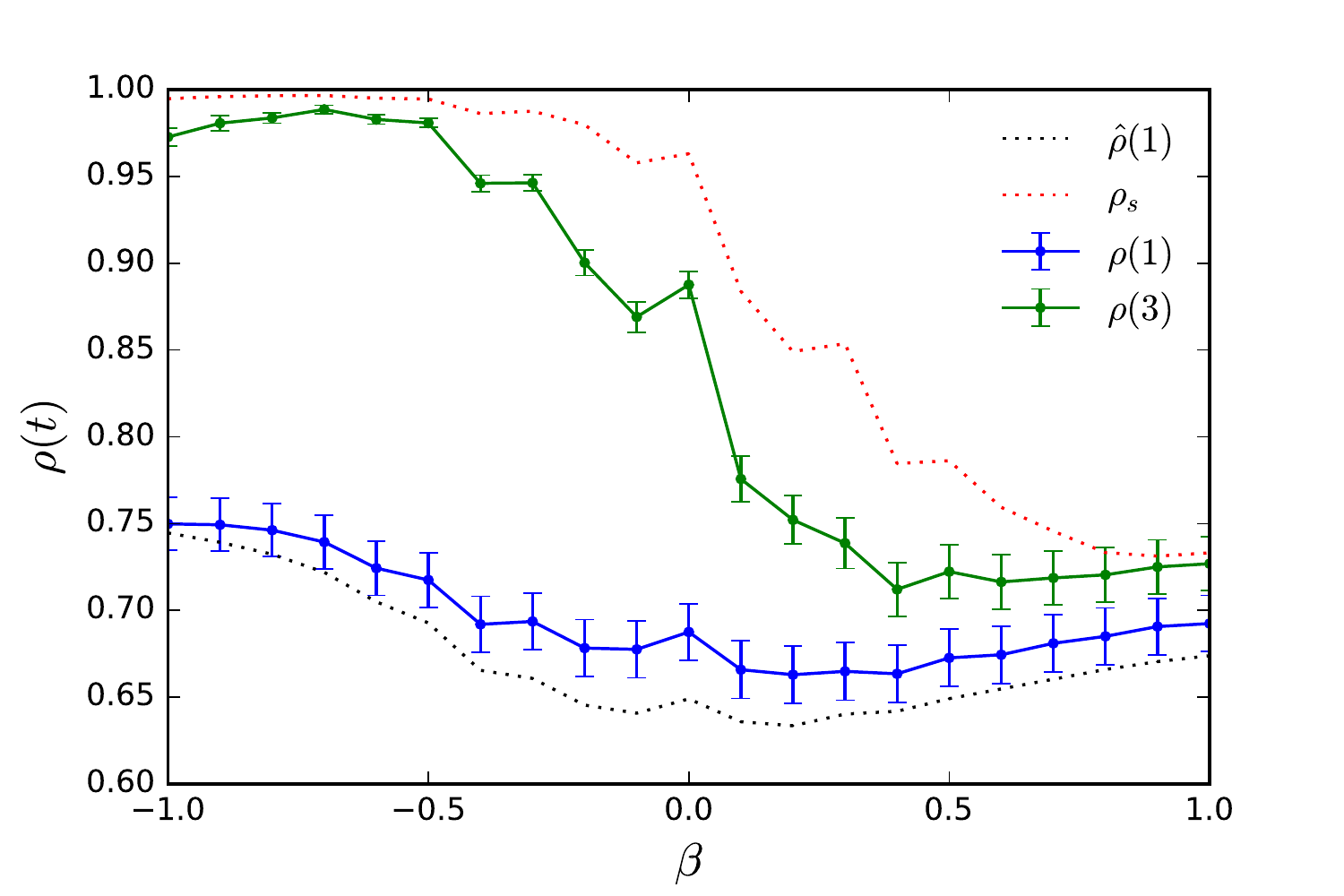}
\\
\end{tabular}
\begin{tabular}{l}
(b)\\
\includegraphics[width=0.94\linewidth]{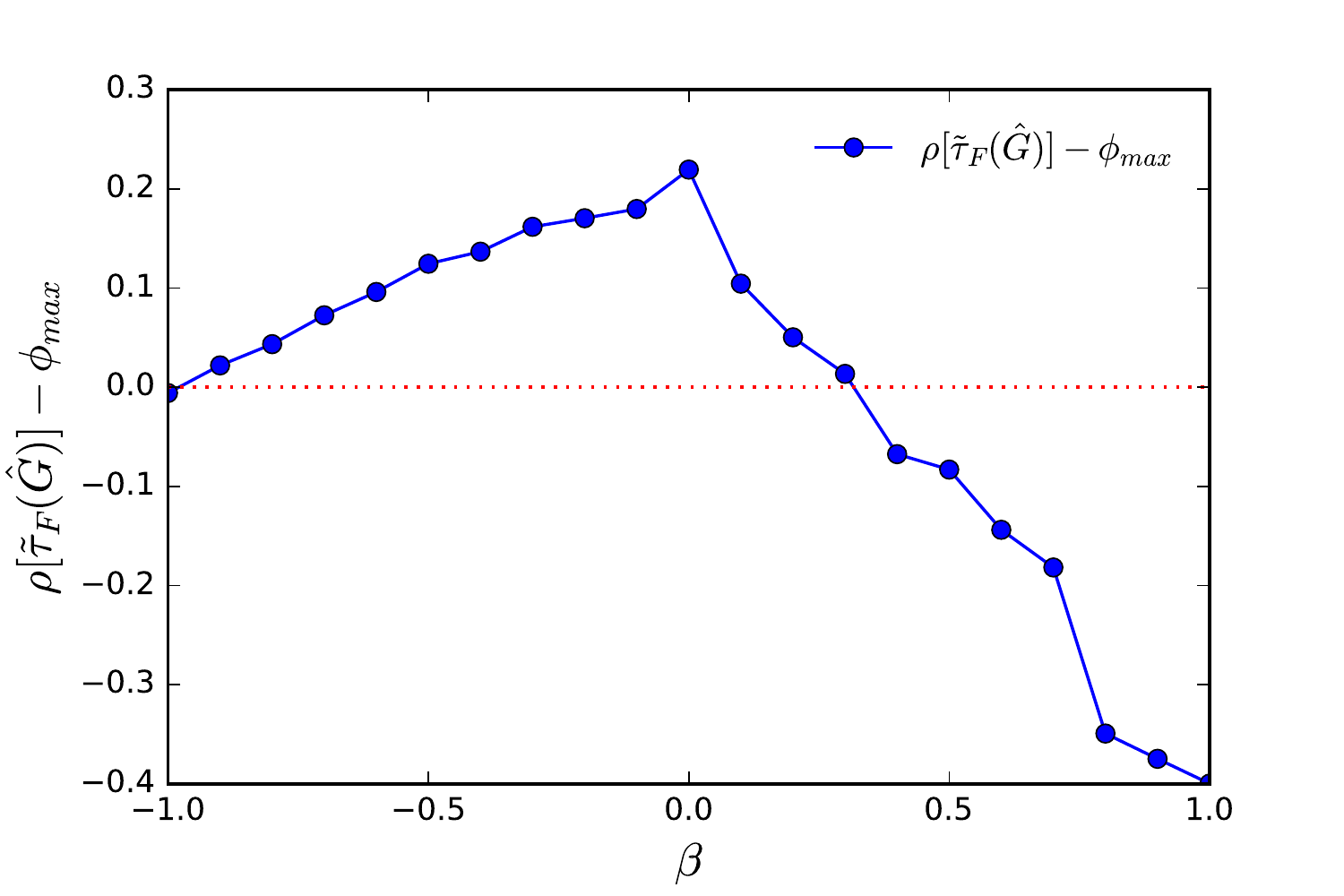}
\\
\end{tabular}

\caption{(Color online) Panel (a) displays the average opinion at $t$ Monte Carlo time step ($\rho(t)$) with $N=800$. The dashed black line is an semi analytic calculation for the initial change (integrating the equations by the Runge-Kutta method). $\rho_s$ is the final average opinion with same system size (error bars show the standard error). Panel (b) shows the difference between average opinion when the initiator group is fixed ($\rho[\tilde{\tau}_F(\hat{G})]$) and maximum threshold ($\phi_\mathrm{max}$). $\beta = 0.4$ is the first point that $\rho[\tilde{\tau}_F(\hat{G})]$ cannot reach to $\phi_\mathrm{max}$ ($N=800$, $n = 500$).}
\label{fig:gap}
\end{figure}

We can obtain $\hat{\rho}(1)$ as the average opinion at $t=1$ by solving Eq.~\eqref{newoi}. This is shown in Fig.~\ref{fig:gap}(a) as black dashed line. The (semi) analytic result in Fig.~\ref{fig:gap})(a) has relatively lower $\hat{\rho}(1)$ around $\beta \in [0.1,0.4]$ over the entire region of $\beta$. The blue line is the numerical result for $\rho(t)$ at $t=1$ ($\rho(1)$), which matches its corresponding analytical calculation well. We can infer from that the region $[0.1\leq \beta \leq 0.4]$ is the region where $P(1\rightarrow0)$ and $P(0\rightarrow1)$ are relatively similar. Thus, in that region, relatively few nodes accept opinion $1$ at $t=1$. With the heterogeneous increment of $1$ at $t=1$, the increment of $\rho(t)$ in $\beta \leq 0$ becomes larger ($t=3$), whereas $\rho(t)$ in $\beta>0$ is barely increased. The asymmetric increment between positive and negative $\beta$ continues until the system reaches the steady state. It means that the innate difference of $\rho(t)$ is magnified by the network structure.

To better understand why the transition of behavior happens at $\beta\approx0.4$ rather than some other $\beta$ value, we used a mean-field approximation. In a mean-field approximation, the opinion of $i$ should be fixed when
\begin{equation}
\rho(t) < \phi_i.  
\label{gap}
\end{equation}
From this, we can see that $\rho(t)$ should be larger than $\phi_\mathrm{max}= \max(\phi_i)$ for all $t$ to spread $1$ to the entire system ($\phi_\mathrm{max}$ can be defined for each $\beta$). Since we want to check whether the Eq.~\eqref{gap} is satisfied or not at the fixation time of $\hat{G}(\beta)$,  Eq.~\eqref{gap} is transformed as $\rho[\tilde{\tau}_{F}(\hat{G})]>\phi_\mathrm{max}$. If it is not satisfied, it means the system with certain $\beta$ is not able to flip the largest threshold node when $\hat{G}(\beta)$ is fixed. 
In Fig.~\ref{fig:gap}(b), it describes the difference of $\phi_\mathrm{max}$ and $\rho[\tilde{\tau}_{F}(\hat{G})]$. It is obvious that $\rho[\tilde{\tau}_{F}(\hat{G})]$ is always larger or similar with $\phi_\mathrm{max}$ when $\beta<0$. For $\beta\simeq 0.4$, $\rho[\tilde{\tau}_{F}(\hat{G})]$ is less than $\phi_\mathrm{max}$. 
For this reason, the most influential nodes are not participating in the spreading process---as seen in Fig.~\ref{fig:fixation}(e)---at an early stage, which hinders change to spread. Indeed, for $\beta\simeq 0.4$ the order of the fixation is reversed so that the influential nodes are the first to fixate. As a result, $\rho_s$ is limited as evident from Fig.~\ref{fig:size} but the opinion within each group ($o_F(i)$) is diverse as can be seen in Fig.~\ref{fig:fixation}(f).  

\begin{figure*}
\begin{tabular}{llll}
(a) & (b) & (c) & (d)\\
\includegraphics[width=0.23\linewidth]{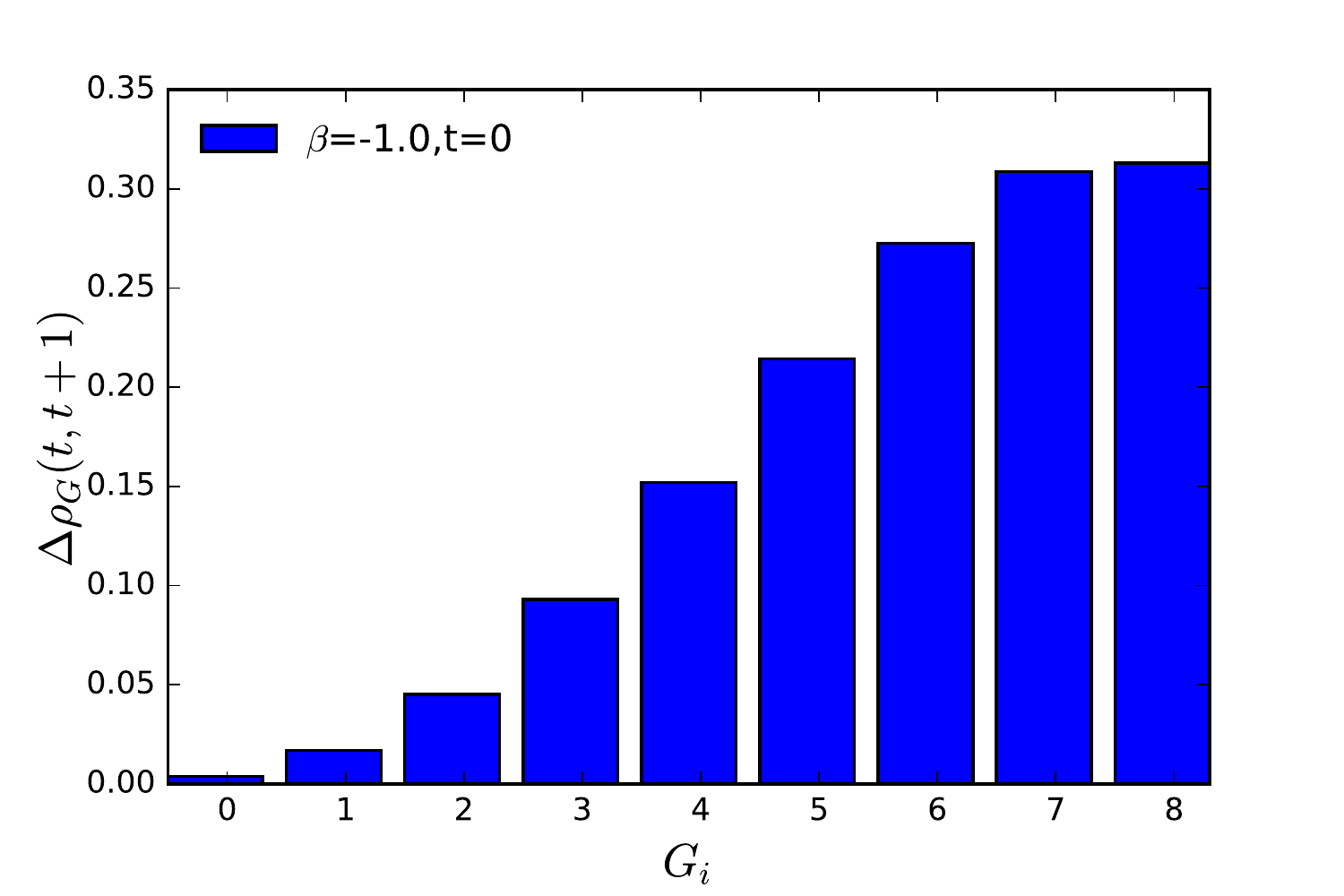} &
\includegraphics[width=0.23\linewidth]{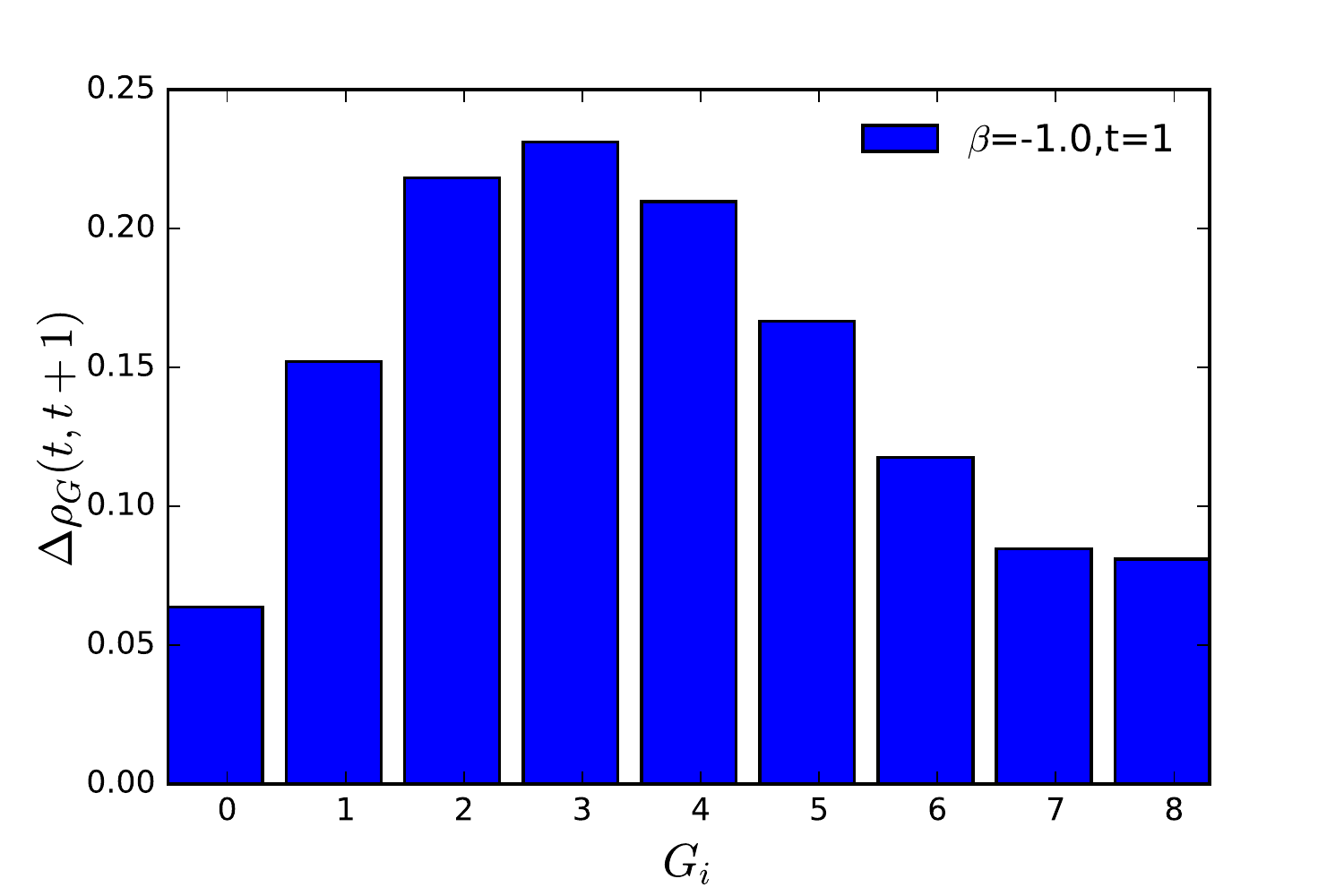}&
\includegraphics[width=0.23\linewidth]{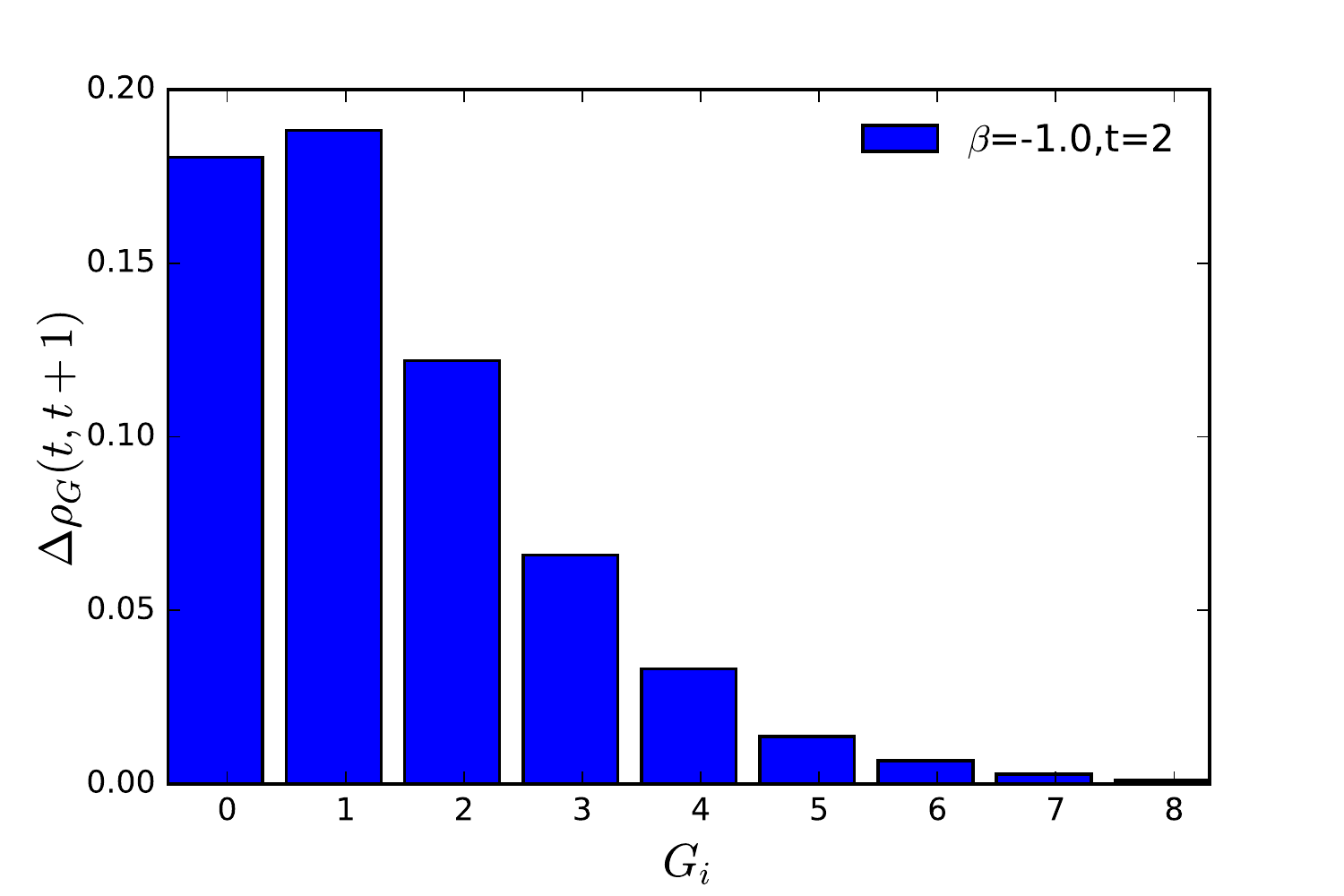}&
\includegraphics[width=0.23\linewidth]{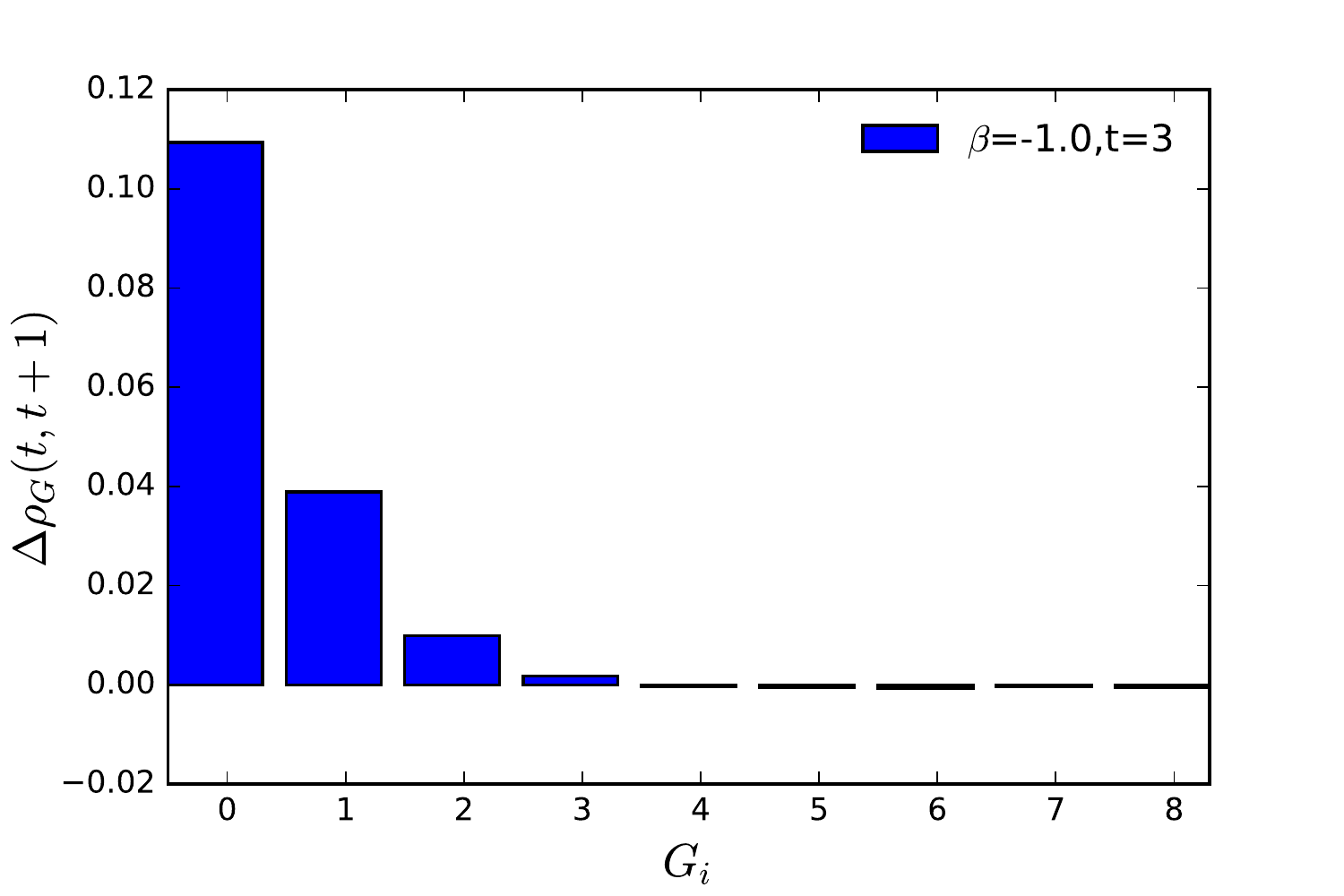}\\
\end{tabular}

\begin{tabular}{llll}
(e) & (f) & (g) & (h) \\
\includegraphics[width=0.23\linewidth]{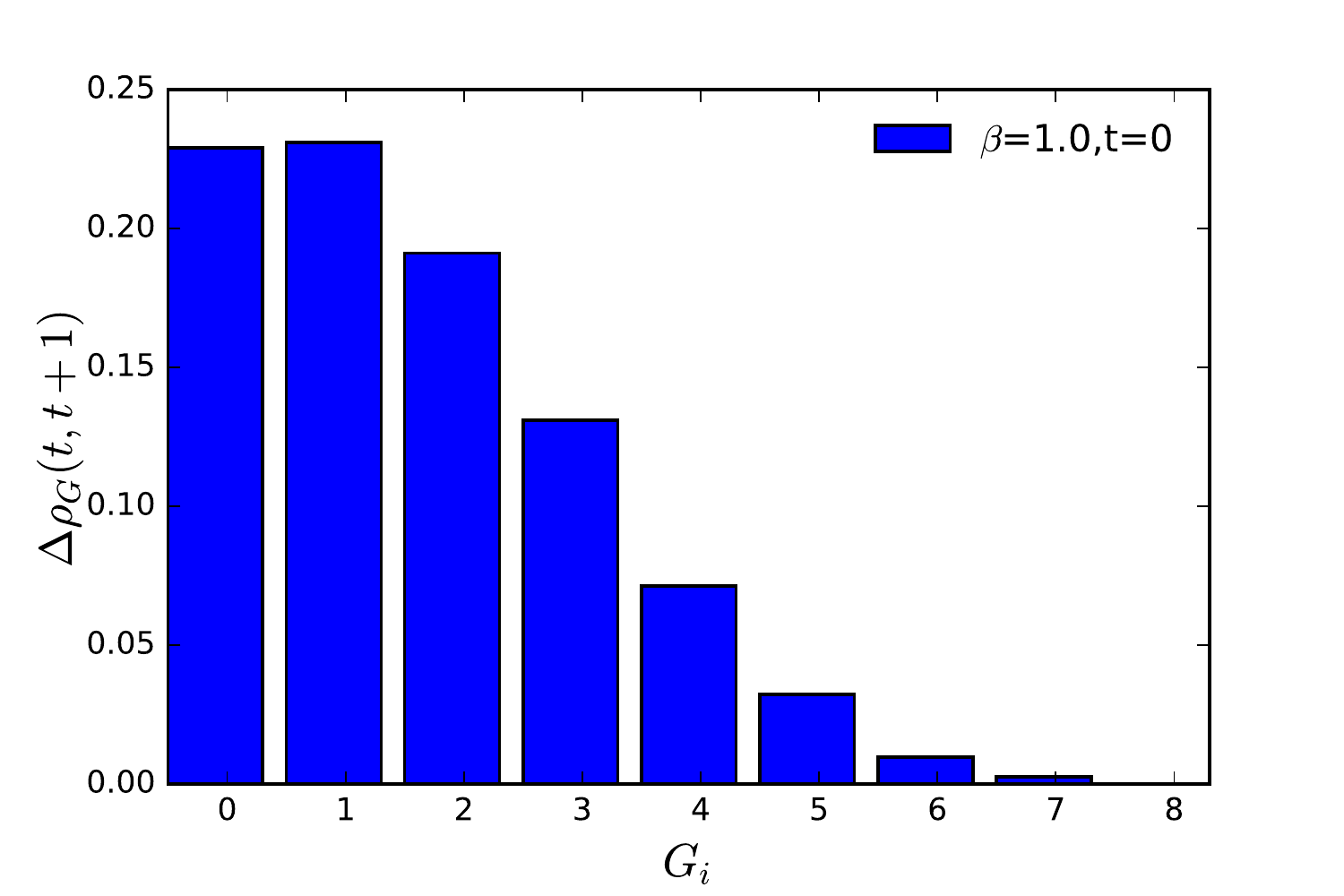} &
\includegraphics[width=0.23\linewidth]{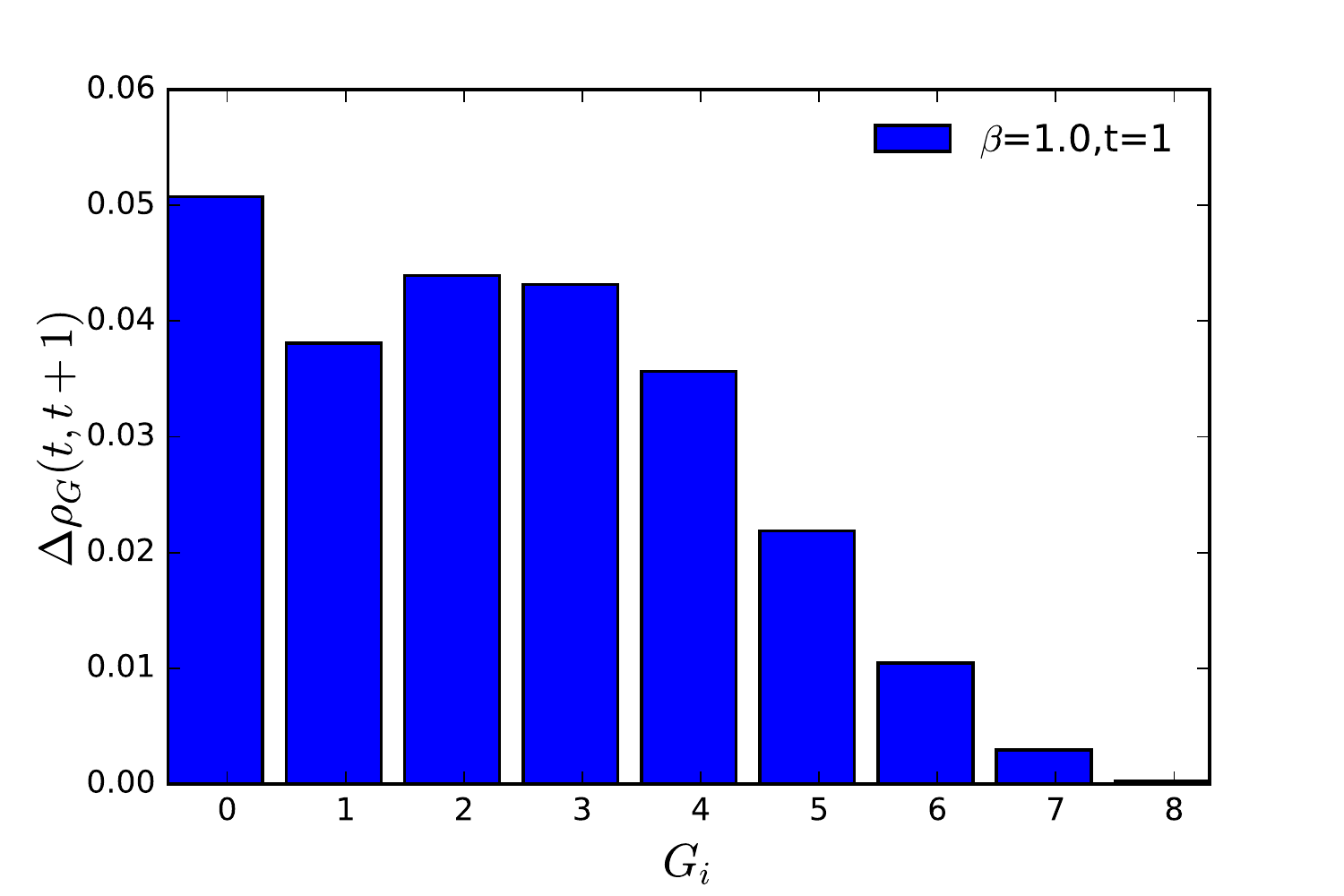}&
\includegraphics[width=0.23\linewidth]{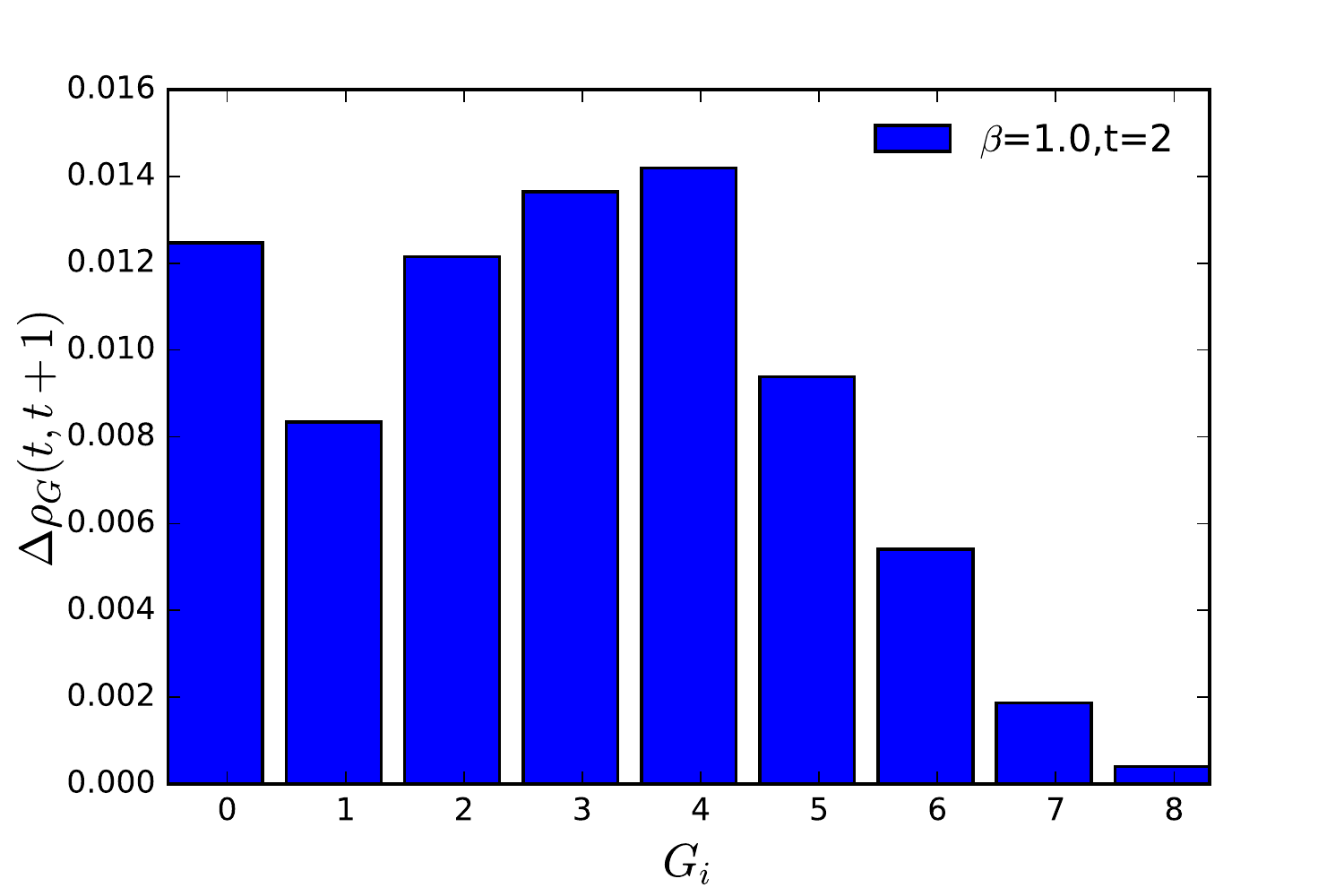}&
\includegraphics[width=0.23\linewidth]{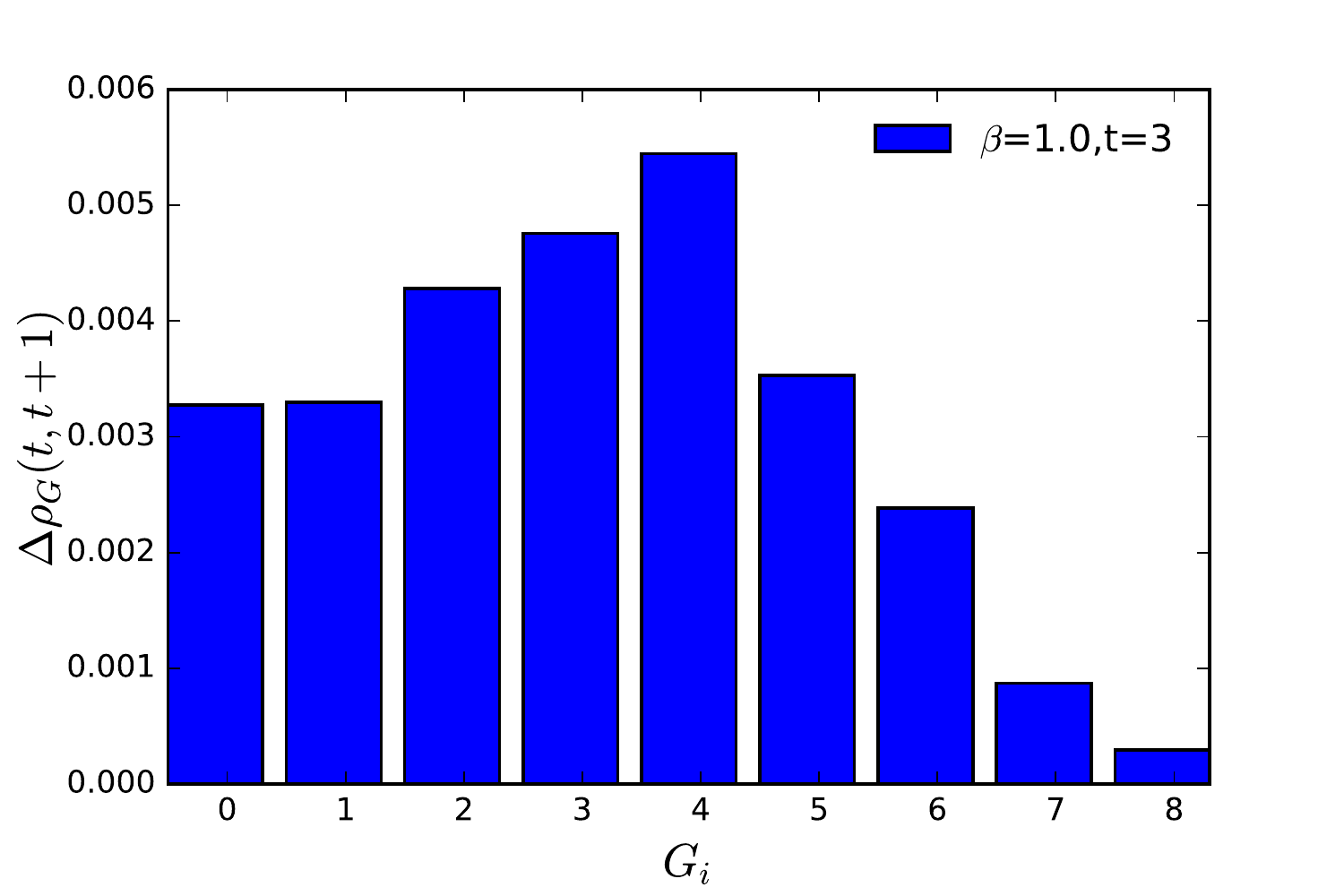}\\
\end{tabular}

\caption{(Color online) The difference on the average opinion of a threshold group by time ($\Delta \rho_{G}(t,t+1)$) for $\beta = -1$ and $1$ ($N=800$, $n = 500$). Negative coupling ($\beta = -1$) shows sequential mechanism and positive coupling ($\beta = 1$) displays segregated mechanism.}
\label{fig:groupdiff}
\end{figure*}

For the very largest $\beta$ value, $\beta=1$, it seems like the spreading once again starts at the influential nodes (Fig.~\ref{fig:fixation}(f)). This case requires a different explanation. To distinguish the difference between the two cases, we measured  $\Delta \rho_{G_q}(t,t+1) = \rho_{G_q}(t+1) - \rho_{G_q}(t)$ for each $q$ (here, $\rho_{G_q}(t)$ is the average opinion of the group $G_q$ at $t$). As representative examples for this study we focus on $\beta = -1,1$. For $\beta = -1$ in Fig.~\ref{fig:groupdiff}(a), $G_8$ and $G_7$ varies most. These variations eventually spreads to $G_0$ (see Fig.~\ref{fig:groupdiff}(b),(c))---the largest threshold group. At $t=3$, the high influence group's changes are saturated so it starts changing opinions. Figs.~\ref{fig:groupdiff}(a),(b),(c) and (d) show that eventually the entire system is participating in the spreading dynamics in a sequential fashion. Even though the step-like $\tilde{\tau}_F(i)$ is very similar to the case of negative coupling, the strongest positive coupling ($\beta = 1$) is not the sequential process. Rather we can call the behavior a segregation mechanism since $G_7$, $G_8$ (which is now the groups of highest influence) are not participating in the spreading from an early stage. As we can see from Fig.~\ref{fig:groupdiff}(f), (g), the initial changes of Fig.~\ref{fig:groupdiff}(e) could not reach to the highest influence groups. In addition, the effect of the high influence nodes exist permanently. As a result, changes circulates within the low influence group. Since changes make  the average opinion of neighbors fluctuate, groups with low threshold (and low influence level) keep fluctuating like $G_0$ through $G_4$ in Fig.~\ref{fig:groupdiff}(e),(f),(g) and (h). The low-influence groups takes more time (high $\tilde{\tau}_F(i)$ as Fig.~\ref{fig:fixation}(f)) to fixate their opinion. In summary, there is a functional segregation in the system that sets the order of fixation. Moreover, the less influential nodes takes more time to set their opinions because of the  fluctuation from high influence nodes they are exposed to. 

\section{Conclusion}
	
We investigate the effect of the coupling between social influence level (out-degree) and a threshold for adoption of opinion change. This problem was recognized in e.g.\ Ref.~\cite{wang}. In our result, there is a clear asymmetric steady state in the negative or mildly positive couplings, on one hand, and positive coupling on the other. There is a phase transition-like behavior around $\beta \simeq 0.4$ separating these two behaviors. For negative $\beta$ values, the effect of the coupling increases the influence of the initiator group so that the new opinion $1$ can reach the entire system. For intermediate $\beta$ values the order of the fixation is reversed, only to be reversed back again at the largest possible $\beta$. However, the reason of the order of fixation is much different in this latter case.
    
The results shares some insights with Ref.~\cite{hidalgo}. In the study, they found the theoretical evidence that the correlation between degree and susceptibility could generate a significant over- or underestimation of the reproductive number (a key quantity in theoretical epidemiology). It also explains that the negative correlation of degree and susceptibility can generate a less vulnerable system  with respect to infectious outbreak. Additionally, the case of positive correlations could work in an opposite way. Even though threshold dynamics and epidemic models have strong difference, we hypothesize that the effects are qualitatively similar.     
    
The obvious continuation of our work would be to establish the phase transition behavior on a firm analytical or computational ground although the latter would maybe be beyond the computational capacity at the moment. Furthermore, it could be possible to see some empirical evidence which will improve an understanding for the spreading dynamics in a society.

\begin{acknowledgments}
The authors were supported by the Basic Science Research Program through the National Research Foundation of Korea (NRF) funded by the Ministry of Education (2016R1D1A1B01007774).
\end{acknowledgments}

\bibliographystyle{apsrev}
\bibliography{thrsh}

\begin{thebibliography}{30}
\expandafter\ifx\csname natexlab\endcsname\relax\def\natexlab#1{#1}\fi
\expandafter\ifx\csname bibnamefont\endcsname\relax
  \def\bibnamefont#1{#1}\fi
\expandafter\ifx\csname bibfnamefont\endcsname\relax
  \def\bibfnamefont#1{#1}\fi
\expandafter\ifx\csname citenamefont\endcsname\relax
  \def\citenamefont#1{#1}\fi
\expandafter\ifx\csname url\endcsname\relax
  \def\url#1{\texttt{#1}}\fi
\expandafter\ifx\csname urlprefix\endcsname\relax\def\urlprefix{URL }\fi
\providecommand{\bibinfo}[2]{#2}
\providecommand{\eprint}[2][]{\url{#2}}

\bibitem[{\citenamefont{Rogers}(2003)}]{rogers}
\bibinfo{author}{\bibfnamefont{E.~M.} \bibnamefont{Rogers}},
  \emph{\bibinfo{title}{Diffusion of Innovations}} (\bibinfo{publisher}{Simon
  \& Schuster}, \bibinfo{address}{New York}, \bibinfo{year}{2003}).

\bibitem[{\citenamefont{Granovetter}(1978)}]{granovetter}
\bibinfo{author}{\bibfnamefont{M.}~\bibnamefont{Granovetter}},
  \bibinfo{journal}{The American Journal of Sociology}
  \textbf{\bibinfo{volume}{83}}, \bibinfo{pages}{1420} (\bibinfo{year}{1978}).

\bibitem[{\citenamefont{Schelling}(1969)}]{schelling}
\bibinfo{author}{\bibfnamefont{T.~C.} \bibnamefont{Schelling}},
  \bibinfo{journal}{The American Economic Review}
  \textbf{\bibinfo{volume}{59}}, \bibinfo{pages}{488} (\bibinfo{year}{1969}).

\bibitem[{\citenamefont{Axelrod}(1997)}]{axelrod}
\bibinfo{author}{\bibfnamefont{R.}~\bibnamefont{Axelrod}}, \bibinfo{journal}{J.
  Conflict Resolut.} \textbf{\bibinfo{volume}{41}}, \bibinfo{pages}{203}
  (\bibinfo{year}{1997}).

\bibitem[{\citenamefont{Watts}(2002)}]{watts}
\bibinfo{author}{\bibfnamefont{D.~J.} \bibnamefont{Watts}},
  \bibinfo{journal}{Proc. Nat. Acad. Sci. USA.} \textbf{\bibinfo{volume}{99}},
  \bibinfo{pages}{5766} (\bibinfo{year}{2002}).

\bibitem[{\citenamefont{Handjani}(1997)}]{handjani}
\bibinfo{author}{\bibfnamefont{S.~J.} \bibnamefont{Handjani}},
  \bibinfo{journal}{Journal of Theoretical Probability}
  \textbf{\bibinfo{volume}{10}}, \bibinfo{pages}{737} (\bibinfo{year}{1997}).

\bibitem[{\citenamefont{Valente}(1996)}]{valente}
\bibinfo{author}{\bibfnamefont{T.~W.} \bibnamefont{Valente}},
  \bibinfo{journal}{Social Networks} \textbf{\bibinfo{volume}{18}},
  \bibinfo{pages}{69} (\bibinfo{year}{1996}).

\bibitem[{\citenamefont{Watts and Dodds}(2007)}]{dodds}
\bibinfo{author}{\bibfnamefont{D.}~\bibnamefont{Watts}} \bibnamefont{and}
  \bibinfo{author}{\bibfnamefont{P.~S.} \bibnamefont{Dodds}},
  \bibinfo{journal}{Journal of Consumer Research}
  \textbf{\bibinfo{volume}{34}}, \bibinfo{pages}{441} (\bibinfo{year}{2007}).

\bibitem[{\citenamefont{Melnik et~al.}(2013)\citenamefont{Melnik, Ward,
  Gleeson, and Porter}}]{melnik}
\bibinfo{author}{\bibfnamefont{S.}~\bibnamefont{Melnik}},
  \bibinfo{author}{\bibfnamefont{J.}~\bibnamefont{Ward}},
  \bibinfo{author}{\bibfnamefont{J.~P.} \bibnamefont{Gleeson}},
  \bibnamefont{and} \bibinfo{author}{\bibfnamefont{M.~A.}
  \bibnamefont{Porter}}, \bibinfo{journal}{Chaos}
  \textbf{\bibinfo{volume}{23}}, \bibinfo{pages}{013124}
  (\bibinfo{year}{2013}).

\bibitem[{\citenamefont{G\'{o}mez et~al.}(2011)\citenamefont{G\'{o}mez, Kappen,
  and Kaltenbrunner}}]{gomez}
\bibinfo{author}{\bibfnamefont{V.}~\bibnamefont{G\'{o}mez}},
  \bibinfo{author}{\bibfnamefont{H.~J.} \bibnamefont{Kappen}},
  \bibnamefont{and}
  \bibinfo{author}{\bibfnamefont{A.}~\bibnamefont{Kaltenbrunner}}, in
  \emph{\bibinfo{booktitle}{Proceedings of the 22nd ACM conference on Hypertext
  and hypermedia}} (\bibinfo{year}{2011}), pp. \bibinfo{pages}{181--190}.

\bibitem[{\citenamefont{Singh et~al.}(2013)\citenamefont{Singh, Sreenivasan,
  Szymanski, and Korniss}}]{singh}
\bibinfo{author}{\bibfnamefont{P.}~\bibnamefont{Singh}},
  \bibinfo{author}{\bibfnamefont{S.}~\bibnamefont{Sreenivasan}},
  \bibinfo{author}{\bibfnamefont{B.~K.} \bibnamefont{Szymanski}},
  \bibnamefont{and} \bibinfo{author}{\bibfnamefont{G.}~\bibnamefont{Korniss}},
  \bibinfo{journal}{Scientific Reports} \textbf{\bibinfo{volume}{3}},
  \bibinfo{pages}{2330} (\bibinfo{year}{2013}).

\bibitem[{\citenamefont{Gleeson and Cahalane}(2007)}]{gleeson2007}
\bibinfo{author}{\bibfnamefont{J.~P.} \bibnamefont{Gleeson}} \bibnamefont{and}
  \bibinfo{author}{\bibfnamefont{D.~J.} \bibnamefont{Cahalane}},
  \bibinfo{journal}{Phys. Rev. E} \textbf{\bibinfo{volume}{75}},
  \bibinfo{pages}{056103} (\bibinfo{year}{2007}).

\bibitem[{\citenamefont{Dodds and Payne}(2009)}]{dodds2}
\bibinfo{author}{\bibfnamefont{P.~S.} \bibnamefont{Dodds}} \bibnamefont{and}
  \bibinfo{author}{\bibfnamefont{J.~L.} \bibnamefont{Payne}},
  \bibinfo{journal}{Phys. Rev. E} \textbf{\bibinfo{volume}{79}},
  \bibinfo{pages}{066115} (\bibinfo{year}{2009}).

\bibitem[{\citenamefont{Whitney}(2010)}]{whitney}
\bibinfo{author}{\bibfnamefont{D.~E.} \bibnamefont{Whitney}},
  \bibinfo{journal}{Phys. Rev. E} \textbf{\bibinfo{volume}{82}},
  \bibinfo{pages}{066110} (\bibinfo{year}{2010}).

\bibitem[{\citenamefont{Gleeson}(2008)}]{gleeson2008}
\bibinfo{author}{\bibfnamefont{J.~P.} \bibnamefont{Gleeson}},
  \bibinfo{journal}{Phys. Rev. E} \textbf{\bibinfo{volume}{77}},
  \bibinfo{pages}{046117} (\bibinfo{year}{2008}).

\bibitem[{\citenamefont{Ya\ifmmode~\breve{g}\else \u{g}\fi{}an and
  Gligor}(2012)}]{osman}
\bibinfo{author}{\bibfnamefont{O.}~\bibnamefont{Ya\ifmmode~\breve{g}\else
  \u{g}\fi{}an}} \bibnamefont{and}
  \bibinfo{author}{\bibfnamefont{V.}~\bibnamefont{Gligor}},
  \bibinfo{journal}{Phys. Rev. E} \textbf{\bibinfo{volume}{86}},
  \bibinfo{pages}{036103} (\bibinfo{year}{2012}).

\bibitem[{\citenamefont{Brummitt et~al.}(2012)\citenamefont{Brummitt, Lee, and
  Goh}}]{charles}
\bibinfo{author}{\bibfnamefont{C.~D.} \bibnamefont{Brummitt}},
  \bibinfo{author}{\bibfnamefont{K.-M.} \bibnamefont{Lee}}, \bibnamefont{and}
  \bibinfo{author}{\bibfnamefont{K.-I.} \bibnamefont{Goh}},
  \bibinfo{journal}{Phys. Rev. E} \textbf{\bibinfo{volume}{85}},
  \bibinfo{pages}{045102} (\bibinfo{year}{2012}).

\bibitem[{\citenamefont{Lee et~al.}(2014)\citenamefont{Lee, Brummitt, and
  Goh}}]{kyumin}
\bibinfo{author}{\bibfnamefont{K.-M.} \bibnamefont{Lee}},
  \bibinfo{author}{\bibfnamefont{C.~D.} \bibnamefont{Brummitt}},
  \bibnamefont{and} \bibinfo{author}{\bibfnamefont{K.-I.} \bibnamefont{Goh}},
  \bibinfo{journal}{Phys. Rev. E} \textbf{\bibinfo{volume}{90}},
  \bibinfo{pages}{062816} (\bibinfo{year}{2014}).

\bibitem[{\citenamefont{Karimi and Holme}(2013)}]{karimi_holme}
\bibinfo{author}{\bibfnamefont{F.}~\bibnamefont{Karimi}} \bibnamefont{and}
  \bibinfo{author}{\bibfnamefont{P.}~\bibnamefont{Holme}},
  \bibinfo{journal}{Physica A} \textbf{\bibinfo{volume}{392}},
  \bibinfo{pages}{3476} (\bibinfo{year}{2013}).

\bibitem[{\citenamefont{Takaguchi et~al.}(2013)\citenamefont{Takaguchi, Masuda,
  and Holme}}]{takaguchi_masuda_holme}
\bibinfo{author}{\bibfnamefont{T.}~\bibnamefont{Takaguchi}},
  \bibinfo{author}{\bibfnamefont{N.}~\bibnamefont{Masuda}}, \bibnamefont{and}
  \bibinfo{author}{\bibfnamefont{P.}~\bibnamefont{Holme}},
  \bibinfo{journal}{PLoS ONE} \textbf{\bibinfo{volume}{8}}, \bibinfo{pages}{1}
  (\bibinfo{year}{2013}).

\bibitem[{\citenamefont{Wang et~al.}(2016)\citenamefont{Wang, Tang, Shu, and
  Wang}}]{wang}
\bibinfo{author}{\bibfnamefont{W.}~\bibnamefont{Wang}},
  \bibinfo{author}{\bibfnamefont{M.}~\bibnamefont{Tang}},
  \bibinfo{author}{\bibfnamefont{P.}~\bibnamefont{Shu}}, \bibnamefont{and}
  \bibinfo{author}{\bibfnamefont{Z.}~\bibnamefont{Wang}}, \bibinfo{journal}{New
  Journal of Physics} \textbf{\bibinfo{volume}{18}}, \bibinfo{pages}{013029}
  (\bibinfo{year}{2016}).

\bibitem[{\citenamefont{Karsai et~al.}(2016)\citenamefont{Karsai, Iñiguez,
  Kikas, Kaski, and Kertész}}]{marton}
\bibinfo{author}{\bibfnamefont{M.}~\bibnamefont{Karsai}},
  \bibinfo{author}{\bibfnamefont{G.}~\bibnamefont{Iñiguez}},
  \bibinfo{author}{\bibfnamefont{R.}~\bibnamefont{Kikas}},
  \bibinfo{author}{\bibfnamefont{K.}~\bibnamefont{Kaski}}, \bibnamefont{and}
  \bibinfo{author}{\bibfnamefont{J.}~\bibnamefont{Kertész}},
  \bibinfo{journal}{Scientific Reports} \textbf{\bibinfo{volume}{6}},
  \bibinfo{pages}{27178} (\bibinfo{year}{2016}).

\bibitem[{\citenamefont{Daniel~Smilkov}(2014)}]{hidalgo}
\bibinfo{author}{\bibfnamefont{L.~K.} \bibnamefont{Daniel~Smilkov},
  \bibfnamefont{Cesar A.~Hidalgo}}, \bibinfo{journal}{Scientific Reports}
  \textbf{\bibinfo{volume}{11}}, \bibinfo{pages}{4795} (\bibinfo{year}{2014}).

\bibitem[{\citenamefont{Miller}(2007)}]{miller}
\bibinfo{author}{\bibfnamefont{J.~C.} \bibnamefont{Miller}},
  \bibinfo{journal}{Phys. Rev. E} \textbf{\bibinfo{volume}{76}},
  \bibinfo{pages}{010101} (\bibinfo{year}{2007}).

\bibitem[{\citenamefont{Kwak et~al.}(2010)\citenamefont{Kwak, Lee, Park, and
  Moon}}]{moon}
\bibinfo{author}{\bibfnamefont{H.}~\bibnamefont{Kwak}},
  \bibinfo{author}{\bibfnamefont{C.}~\bibnamefont{Lee}},
  \bibinfo{author}{\bibfnamefont{H.}~\bibnamefont{Park}}, \bibnamefont{and}
  \bibinfo{author}{\bibfnamefont{S.}~\bibnamefont{Moon}}, in
  \emph{\bibinfo{booktitle}{{P}roceedings of the 19th {I}nternational {W}orld
  {W}ide {W}eb {C}onference}} (\bibinfo{year}{2010}).

\bibitem[{\citenamefont{Newman}(2005)}]{mejn:powerlaw}
\bibinfo{author}{\bibfnamefont{M.~E.~J.} \bibnamefont{Newman}},
  \bibinfo{journal}{Contemporary Physics} \textbf{\bibinfo{volume}{46}},
  \bibinfo{pages}{323} (\bibinfo{year}{2005}).

\bibitem[{\citenamefont{Mirta~Galesic}(2012)}]{galesic}
\bibinfo{author}{\bibfnamefont{J.~R.} \bibnamefont{Mirta~Galesic},
  \bibfnamefont{Henrik~Olsson}}, \bibinfo{journal}{Psychological Science}
  \textbf{\bibinfo{volume}{23}}, \bibinfo{pages}{1515} (\bibinfo{year}{2012}).

\bibitem[{\citenamefont{Dunbar}(1992)}]{dunbar}
\bibinfo{author}{\bibfnamefont{R.~I.~M.} \bibnamefont{Dunbar}},
  \bibinfo{journal}{Journal of Human Evolution} \textbf{\bibinfo{volume}{22}},
  \bibinfo{pages}{469 } (\bibinfo{year}{1992}).

\bibitem[{\citenamefont{Newman}(2010)}]{newman:book}
\bibinfo{author}{\bibfnamefont{M.~E.~J.} \bibnamefont{Newman}},
  \emph{\bibinfo{title}{Networks: An Introduction}} (\bibinfo{publisher}{Oxford
  University Press}, \bibinfo{address}{Oxford}, \bibinfo{year}{2010}).

\bibitem[{\citenamefont{Chung and Lu}(2002)}]{chunglu}
\bibinfo{author}{\bibfnamefont{F.}~\bibnamefont{Chung}} \bibnamefont{and}
  \bibinfo{author}{\bibfnamefont{L.}~\bibnamefont{Lu}}, \bibinfo{journal}{Proc.
  Nat. Acad. Sci. USA.} \textbf{\bibinfo{volume}{99}}, \bibinfo{pages}{15879}
  (\bibinfo{year}{2002}).

\end{thebibliography}

\end{document}